\documentclass{bmcart}

\usepackage{amsthm,amsmath}
\usepackage{ulem}
\usepackage[utf8]{inputenc} %unicode support
\usepackage{graphicx, rotating}
\usepackage{caption}
\usepackage{subcaption}
\usepackage{tabularx}
\usepackage{makecell}
\usepackage{pbox}
\usepackage{booktabs}
\usepackage{upgreek}
\usepackage{pdfpages}
\usepackage{multirow}
\usepackage[colorlinks = true,
            linkcolor = black,
            citecolor = blue,
            anchorcolor = blue]{hyperref}

\usepackage{diagbox}
\usepackage[colorinlistoftodos]{todonotes}
\usepackage{amsmath}
\usepackage{bm}% bold math
\usepackage{siunitx}
\usepackage{wrapfig}

\startlocaldefs
\endlocaldefs

\begin{document}

%%% Start of article front matter
\begin{frontmatter}

%\begin{fmbox}
\dochead{Research}


\title{Design of an enhanced mechanism for a new Kibble balance directly traceable to the quantum SI}

%%%%%%%%%%%%%%%%%%%%%%%%%%%%%%%%%%%%%%%%%%%%%%
%%                                          %%
%% Enter the authors here                   %%
%%                                          %%
%% Specify information, if available,       %%
%% in the form:                             %%
%%   <key>={<id1>,<id2>}                    %%
%%   <key>=                                 %%
%% Comment or delete the keys which are     %%
%% not used. Repeat \author command as much %%
%% as required.                             %%
%%                                          %%
%%%%%%%%%%%%%%%%%%%%%%%%%%%%%%%%%%%%%%%%%%%%%%

\author[]{\inits{L.}\fnm{Lorenz} \snm{Keck}}
\author[]{\inits{F.}\fnm{Frank} \snm{Seifert}}
\author[]{\inits{D.}\fnm{David} \snm{Newell}}
\author[]{\inits{S.}\fnm{Stephan} \snm{Schlamminger}}
\author[]{\inits{R.}\fnm{Ren\'e} \snm{Theska}}
\author[]{\inits{D.}\fnm{Darine} \snm{Haddad}}

\begin{abstractbox}

\begin{abstract} % abstract
The ``Quantum Electro-Mechanical Metrology Suite'' (QEMMS) is being designed and built at the National Institute of Standards and Technology. It includes a Kibble balance, a graphene quantum Hall resistance array and a Josephson voltage system, so that it is a new primary standard for the unit of mass, the kilogram, directly traceable to the International System of Units (SI) based on quantum constants. We are targeting a measurement range of \SI{10}{\gram} to \SI{200}{\gram} and optimize the design for a relative combined uncertainty of \SI{2e-8}{} for masses of \SI{100}{\gram}. QEMMS will be developed as an open hardware and software design. In this article, we focus on the design of an enhanced moving and weighing mechanism for the QEMMS based on flexure pivots.

%\parttitle{First part title} %if any
%Text for this section.

%\parttitle{Second part title} %if any
%Text for this section.
\end{abstract}

%%%%%%%%%%%%%%%%%%%%%%%%%%%%%%%%%%%%%%%%%%%%%%
%%                                          %%
%% The keywords begin here                  %%
%%                                          %%
%% Put each keyword in separate \kwd{}.     %%
%%                                          %%
%%%%%%%%%%%%%%%%%%%%%%%%%%%%%%%%%%%%%%%%%%%%%%


% MSC classifications codes, if any
%\begin{keyword}[class=AMS]
%\kwd[Primary ]{}
%\kwd{}
%\kwd[; secondary ]{}
%\end{keyword}

\end{abstractbox}
%
%\end{fmbox}% uncomment this for two column layout

\end{frontmatter}

%%%%%%%%%%%%%%%%%%%%%%%%%%%%%%%%%%%%%%%%%%%%%%%%
%%                                            %%
%% The Main Body begins here                  %%
%%                                            %%
%% Please refer to the instructions for       %%
%% authors on:                                %%
%% https://www.biomedcentral.com/getpublished %%
%% and include the section headings           %%
%% accordingly for your article type.         %%
%%                                            %%
%% See the Results and Discussion section     %%
%% for details on how to create sub-sections  %%
%%                                            %%
%% use \cite{...} to cite references          %%
%%  \cite{koon} and                           %%
%%  \cite{oreg,khar,zvai,xjon,schn,pond}      %%
%%                                            %%
%%%%%%%%%%%%%%%%%%%%%%%%%%%%%%%%%%%%%%%%%%%%%%%%

%%%%%%%%%%%%%%%%%%%%%%%%% start of article main body
% <put your article body there>

%%%%%%%%%%%%%%%%
%% Background %%
%%
%\section*{Content}
%Text and results for this section, as per the individual journal's instructions for authors.

\section*{Introduction}
The Kibble balance is the instrument for the precise primary realization of the unit of mass in the International System of Units (SI) based on electro-mechanical metrology principles and quantum physics. Due to the design of a new version of the Kibble balance at the National Institute of Standards and Technology (NIST) as part of the so-called ``Quantum Electro-Mechanical Metrology Suite", QEMMS, this article describes the components integrated in QEMMS as well as the design of the new flexure-based moving and weighing mechanism.

\section*{The Kibble balance}
The measuring principle of the Kibble balance was invented in 1976 by the metrologist Dr. Bryan Kibble at the National Physical Laboratory of the United Kingdom~\cite{Kibble1976}. It was one of the main experiments contributing to the redefinition of the SI in the year 2018~\cite{BIPM2019, Tiesinga2021}. The redefinition of the unit of mass, the kilogram, based on the fixed values of three constants of nature -- the Planck's constant, $h$, the speed of light, $c$, and the hyperfine transition frequency of Caesium, $\Delta\nu_{\mbox{Cs}}$ -- provides a means to realize mass through electrical metrology using absolute precision balances without the need to compare to a physical object.

\begin{figure}[h!]
    \centering
    \includegraphics[width=0.4\columnwidth]{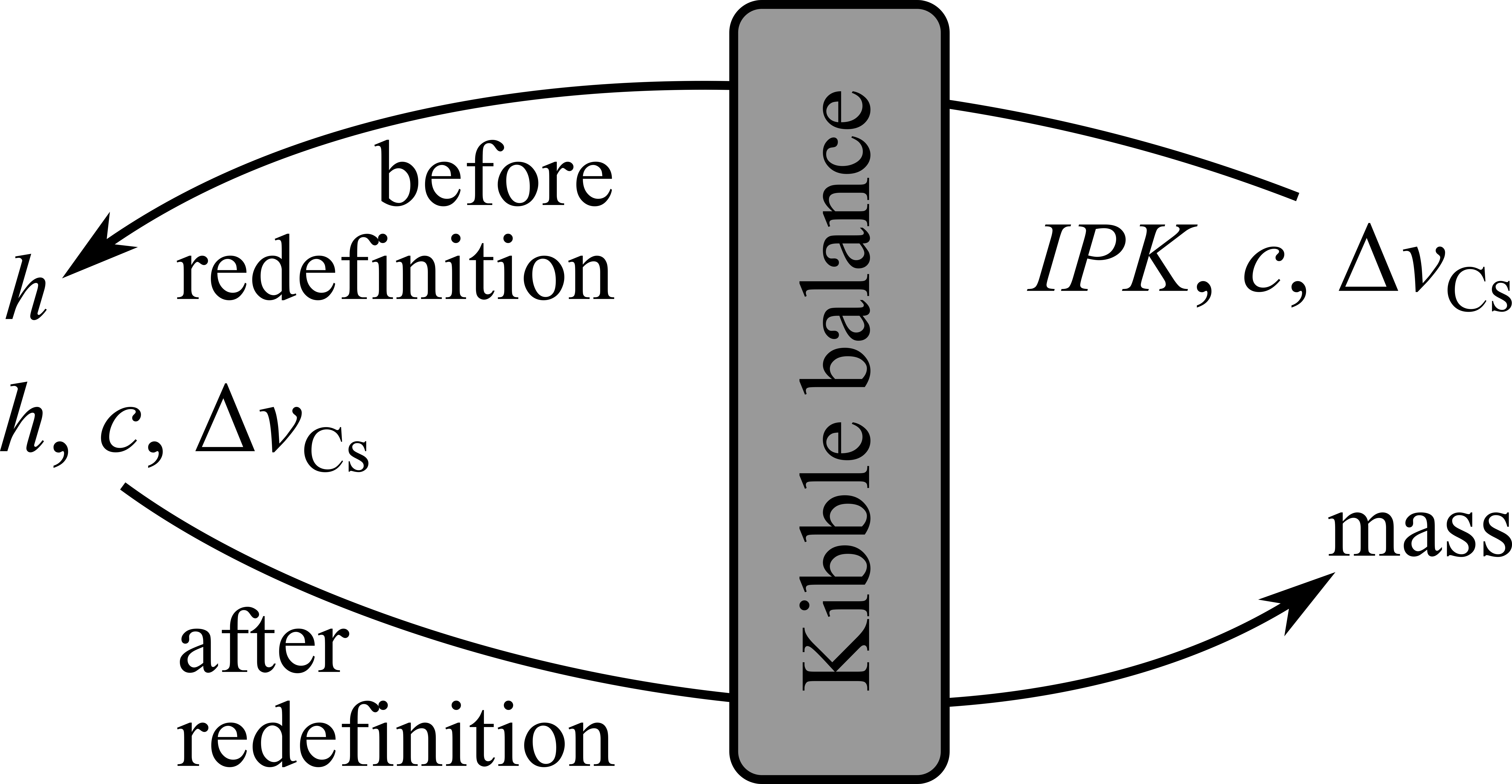}
    \caption{The relation between the IPK, natural constants and the unit of mass in the Kibble balance in traditional and reversed operation.}
    \label{fig:KibbAndH}
\end{figure}

Until 2018, only the values for $c$ and $\Delta\nu_{\mbox{Cs}}$ were fixed, and the International Prototype Kilogram (IPK) was the last physical artifact defining one of the seven SI base units. Using the Kibble balance principle and decades of hard work, scientists were able to measure and define a numerical value for $h$ based on the mass of the IPK, the speed of light and the hyperfine transition frequency of Caesium. As a result, the Kibble balance can be operated in a reversed fashion and realize virtually any macroscopic mass value directly from the defined value of $h$. A visualization of this relation is shown in figure~\ref{fig:KibbAndH}. 

The Kibble balance uses two modes of operation for an absolute mass measurement.
In the first one, the weighing mode, the gravitational force of a test mass is directly compensated by a counter force such that the test mass stays at a defined null position. The counter force is provided by a magnet-coil system and can be controlled by adjusting the current through the coil.

\begin{figure*}[t]
  \centering
  \begin{minipage}[b]{0.45\columnwidth}
    \includegraphics[width=.99\columnwidth]{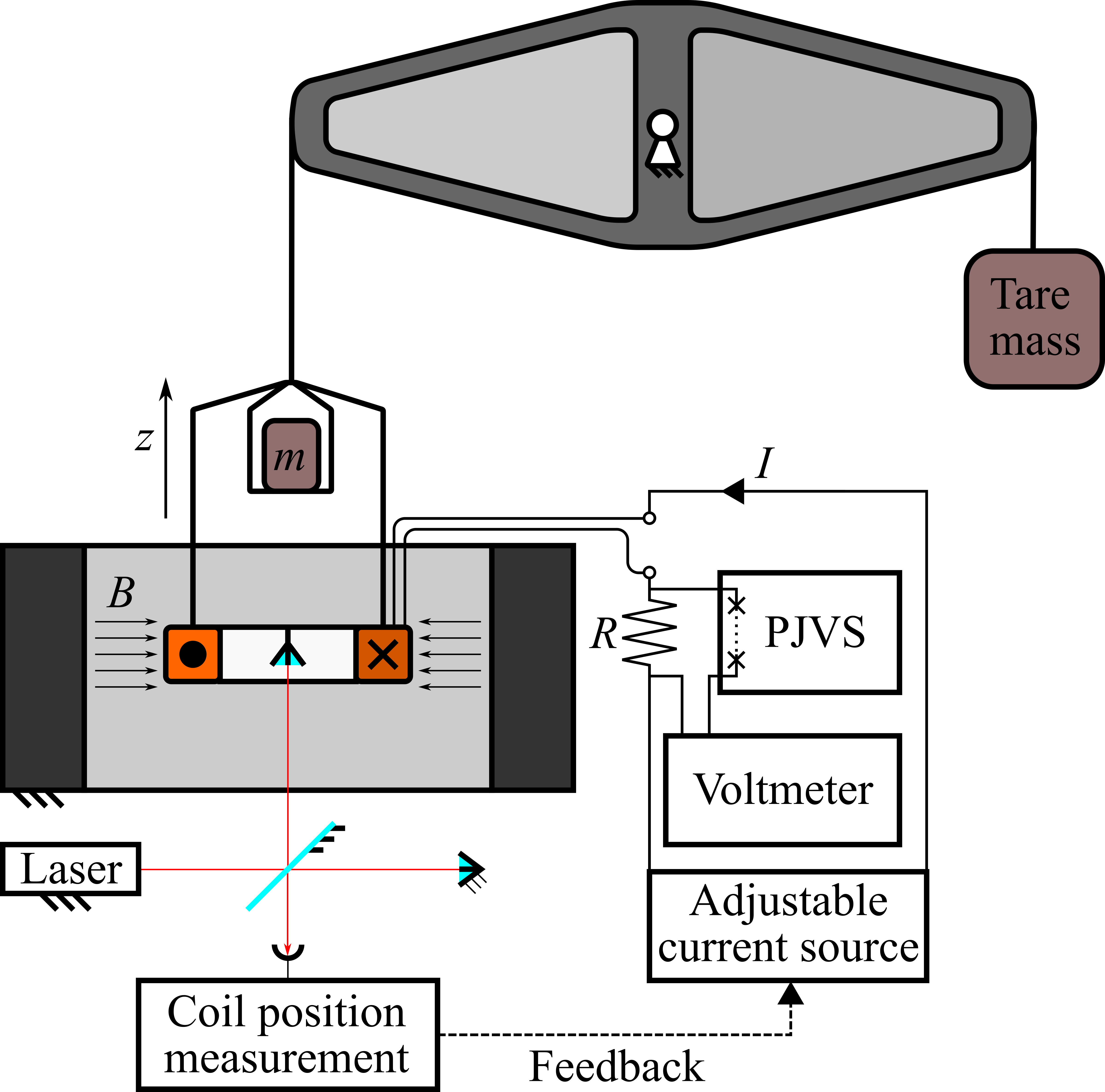}
    \caption{The principle of operation in the Kibble balance in the weighing mode. The gravitational force of a test mass $m$ is compared with the magnetic force produced by the magnet-coil system to keep the test mass in a defined, feedback-controlled position.}
    \label{fig:KibbOperationWeigh}
  \end{minipage}
  \hfill
  \begin{minipage}[b]{0.45\columnwidth}
    \includegraphics[width=.99\columnwidth]{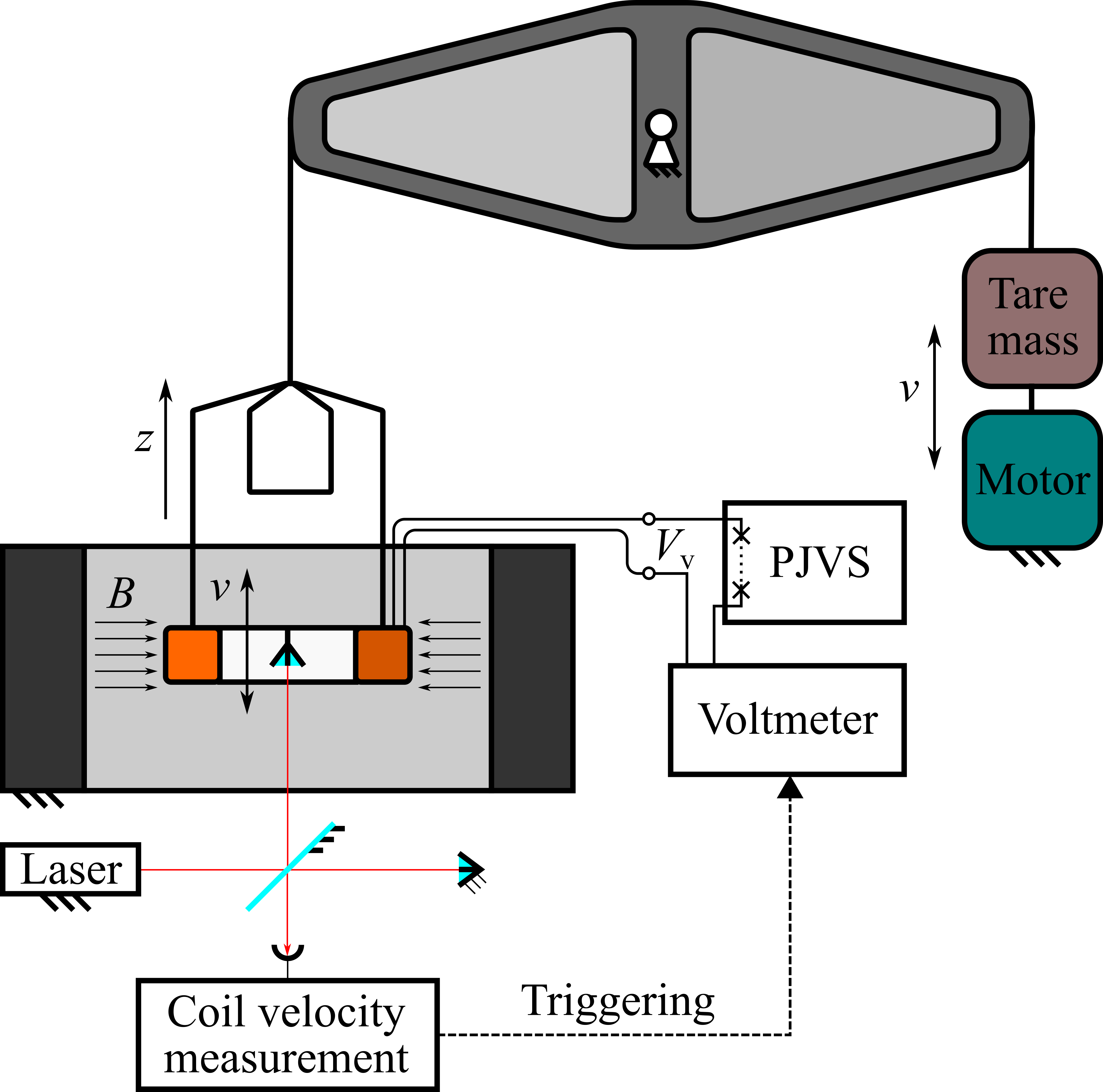}
    \caption{Schematics for the principle of operation in the Kibble balance in the velocity mode. The coil is moved by a motor through the magnetic field of the magnet with vertical velocity $v$, and a voltage $V$ is induced in the coil. The voltage measurement is synchronized with the position measurement of the coil using a time counter.}
    \label{fig:KibbOperationVel}
  \end{minipage}
\end{figure*}

The equation for this mode of operation is
\begin{equation}
    - m \, g = N \, \frac{\partial \Phi}{\partial z} \, I,
    \label{eq:WeighMode}
\end{equation}

where $m$ is the mass of the test mass, $g$ the local gravitational acceleration, $N$ the number of turns of the wire in the coil, $\Phi$ the magnetic flux through the coil, $z$ the position of the coil along the vertical direction, and $I$ the electric current through the coil.

Since the magnetic flux gradient $N \, \partial \Phi / \partial z$ is very hard to obtain with sufficient uncertainty, a second mode of operation, the velocity mode, allows for direct measurement of the $N \, \partial \Phi / \partial z$ with the required precision. Here, the coil is moved up and down in the magnetic field of the magnet such that a voltage is induced in the coil. Derived from Faraday's induction law it yields,

\begin{equation}
    V = -N \, \frac{\partial \Phi}{\partial z} v,
    \label{eq:VelMode}
\end{equation}

where $V$ is the voltage induced in the coil in the velocity mode, and $v$ is the vertical velocity of the coil with respect to the magnet. The velocity in high precision Kibble balances is usually measured with an interferometer and vacuum operation of the balance creates an environment with a stable index of refraction. Solving both equation~(\ref{eq:WeighMode}), and equation~(\ref{eq:VelMode}) for $N \, \partial \Phi / \partial z$ and equating them yields

\begin{equation}
    m \, g \, v = I \, V.
    \label{eq:WattEq}
\end{equation}

On the left hand side in equation~(\ref{eq:WattEq}), we see the expression for mechanical power, and on the right hand side, the one for electrical power. The original name of the Kibble balance, ``watt balance", is a reflection of the power balance principle described here.\\

A direct measurement of mass traceable to the new SI as in the QEMMS would not be possible without two quantum effects: The Josephson effect and the quantum Hall effect. The electric current $I$ through the coil in the weighing mode is usually measured according to Ohm's law by monitoring a voltage drop over a traditional resistor $R$ with very precisely known value (see figure~\ref{fig:KibbOperationWeigh}), which is calibrated against a quantum Hall resistor (QHR) standard $R_\mathrm{H} = h/(i e^{2})$ where $i$ is the Landau level filling factor~\cite{Delahaye2003,Jeckelmann2001} and $e$ is the elementary charge. A Programmable Josephson Voltage System (PJVS) is a frequency to voltage converter, that is, $V=\nu/K_J$ where $K_J=2e/h$~\cite{Hamilton2000} and $\nu$ is a frequency. The PJVS provides a primary realization for voltage and is put parallel to the resistor in the electric circuit. It is used to create a voltage nominally equal to the drop over the resistor and a calibrated voltmeter monitors the remaining potential difference in the circuit. The voltage, hence, appears in the weighing mode, and the velocity mode. Adding the Josephson and quantum Hall effect to equation~\ref{eq:WattEq} yields the link of mass with Planck's constant established with the Kibble balance:

\begin{equation}
    m = \frac{i \, n^2 \, \nu^2}{4 \, v \, g} \, h.
    \label{eq:Massh}
\end{equation}

Furthermore, the velocity and the local gravitational acceleration in equation~\ref{eq:Massh} are measured using primary standards of length and time based on the definition of the second and meter traceable to $\Delta\nu_{\mbox{Cs}}$ and $c$, respectively. 

More detailed information on the link of the quantum effects with the Kibble balance equation can be found in~\cite{Haddad2016b, Robinson2016a}.

\section*{The QEMMS}

There are multiple Kibble balance experiments around the globe. The design of each Kibble balance is unique, however, one feature seems to be common to all:
The resistor in the electrical circuit for the weighing mode (see figure~\ref{fig:KibbOperationWeigh}) is calibrated using transfer standards against a QHR as primary standard for resistance. QEMMS will overcome the inconvenience of external resistor calibration and thus eliminate the resistance calibration uncertainty by implementing a QHR directly in the Kibble balance's electrical circuit. This QHR is based on the quantization effect of monolayer graphene~\cite{Kruskopf2016}. Among the advantages offered by graphene-based QHR standards is their ability to maintain the fully quantized state at higher currents and temperature. NIST researchers fabricated thirteen single element graphene QHR standards in parallel and operate at the Landau level $i = 2$ plateau, giving a nominal value of $R_\mathrm{K}/26 \approx \SI{992.8}{\ohm}$~\cite{Panna2021}. Even though the graphene QHR array in Panna \textit{et al.} show experiments with maximum currents up to \SI{0.3}{\milli \ampere}, measurements have already been published showing single Hall bar graphene devices maintaining quantization for currents up to \SI{0.7}{\milli \ampere}~\cite{Rigosi2019}. Furthermore, creating arrays with higher currents is subject to extensive research efforts at NIST. In context of QEMMS, these allow for a measurement of masses up to \SI{200}{\gram}, depending on the design of the coil in the magnetic field~\cite{Marangoni2020}.

Hence, this will be the first integrated instrument in the world to feature all three: a Kibble balance for the actual mass measurement, a PJVS, and a primary reference resistor directly implemented in the instrument, which provides an outlook for improved overall measurement uncertainty for masses. The entire QEMMS will fit in one room of the size of approximately $\SI{4}{\meter} \times \SI{5}{\meter}$ (width $\times$ length). Through the direct link to all three natural constants $h$, $c$ and $\Delta\nu_{\mbox{Cs}}$, it is providing direct SI realization for the units of mass, time, length, electrical resistance, voltage, and current. Thus, the QEMMS can be seen as a metrology institute in one single experiment. 

As opposed to traditional Kibble balances~\cite{Robinson2016a} which were designed and optimized to measure \SI{1}{\kilogram}, the QEMMS is being built to cover a mass range of $\SI{10}{\gram}$ to $\SI{200}{\gram}$. We optimize the design specifically targeting a relative combined measurement uncertainty of \SI{2e-8}{} at nominal mass value of \SI{100}{\gram}. Here, we want to take advantage of the direct implementation of the new quantum SI in the instrument for mass scaling. A factor of 20 in the total mass range has proven convenient for systematic studies on the balance, which motivates the covered range around the nominal value. Furthermore, smaller nominal mass values allow for a more robust and compact design, which makes the technology more accessible for other metrology institutes as well.

Since a QHR has been designed for this purpose~\cite{Panna2021}, we put the focus on the redesign and optimization of the Kibble balance. Each individual subsystem of the Kibble balance is being investigated regarding new opportunities for improvement. The permanent magnet system~\cite{Marangoni2020} and a vacuum chamber~\cite{Marangoni2020b} have already been acquired. In the current design step, improvement of the mechanical components in the balance, especially the mechanism, is of key interest since the state of the art seems to be one of the limiting factors for the overall balance uncertainty. Hence, the design of the QEMMS mechanism is the focus of the following chapters.

\subsection*{Requirements to the QEMMS mechanism}\label{Mechanisms}

The mechanism in the QEMMS has two functions: It provides suspension and defines the trajectory of moving components. The former entails suspending all functional components of the balance such as the mass pan, the coil, a counterweight for balancing, the coil suspension and optics for the interferometer. The latter requires the mechanism to move the coil in a vertical trajectory through the magnet in velocity mode. We are seeking to integrate these two main functions in the same mechanism so that one mechanical system is used to perform both modes of operation in the QEMMS.

\begin{table}[ht]
    \begin{center}
            \caption{Requirements for the design of the QEMMS mechanism}
            \begin{tabular}{ l  c  c }
                \toprule
                \textbf{Requirement} & \textbf{Value} & \textbf{Unit} \\
                \hline
                
                Coil travel in velocity mode & $\pm 30$ & \SI{}{\milli \meter} \\
                
                Stiffness$^1$ & $\leq 0.01$ & \SI{}{\newton \per \meter} \\
                
                Linear guiding of coil & -- & -- \\
                
                Guiding imperfection of coil & $\leq 10$ & \SI{}{\micro \meter} \\
                
                Hysteretic force & $\leq 0.1$ & \SI{}{\micro \gram} \\
                
                Carrying capacity & $\approx 15$ & \SI{}{\kilo \gram} \\
                
                Installation space$^2$ & $\leq 500$ & \SI{}{\milli \meter} \\
                \bottomrule
            \end{tabular}\\
            \label{tab:requirements}
    \end{center}

    \footnotesize{$^1$ The stiffness should be constant along the travel in the velocity mode, and over time. It should not change with, e.g., temperature fluctuation.\\
    $^2$ This indicates the maximum horizontal dimension.}
\end{table}

A set of important requirements pre-define a successful mechanism design. The most important aspects that can directly be impacted by conceptual decisions in an early state of design are listed in table~\ref{tab:requirements}.

In general, it is important to keep all uncertainty contributions of the mechanical system to the measurement sufficiently small that their combination does not exceed the targeted uncertainty. For the QEMMS we define this as on the order of parts in $10^{9}$. In the following, each requirement from table~\ref{tab:requirements} is explained in more detail regarding the background of its origin and importance. 

\subsubsection*{Coil travel in the velocity mode}

For a measurement of $N \, \partial \Phi / \partial z$ in the velocity mode, the coil will be moved up and down in the magnetic field of the permanent magnet for the QEMMS. At first, it is accelerated to a velocity of $v_\mathrm{z}=\SI{2}{\milli\meter \per \second}$. Then it will travel with constant speed for \SI{40}{\milli \meter}. Finally, the coil is decelerated to a stop at the end of each sweep. The permanent magnet for the QEMMS has been designed and built to provide uniform $\partial \Phi / \partial z$ profile over the course of the travel with constant speed. This area is referred to as the precision air gap.~\cite{Marangoni2020}

Hence, the travel range of the coil is given by the length of the precision air gap plus additional travel for acceleration and deceleration of the coil. A conservative estimate of the total travel length based on experience with previous Kibble balance designs is $\SI{60}{\milli \meter}$. We will use this number as an input parameter for the mechanism design of QEMMS.

\subsubsection*{Stiffness}

To keep measurement errors at the necessary minimum and to achieve the required resolution, the balance mechanism needs to have a low stiffness. More importantly, we want the stiffness to be constant in time and over all possible coil positions for the following three reasons:\\

(a) The noise in the balance position feedback converts to force noise by multiplying the position noise with the stiffness of the mechanism. Hence, a low stiffness produces a small force noise at a stable position which reduces averaging time.
Since QEMMS will be optimized to measure \SI{100}{\gram} mass artifacts with an overall relative standard uncertainty of \SI{2e-8}{}, we are aiming for a resolution of the balance of \SI{1e-9}{} on \SI{100}{\gram}. This yields an absolute mass resolution of \SI{0.1}{\micro \gram}, which is roughly equivalent to a force resolution of \SI{1}{\nano \newton}. We assume we can average the position feedback uncertainty down to \SI{100}{\nano \meter}, which means that the mechanism needs to have a stiffness of $\leq \SI{0.01}{\newton \per\meter}$ at the weighing position to be within the resolution requirements.\\

(b) A constant stiffness value along the coil travel is important for an accurate and stable feedback control during the velocity mode. A higher order stiffness term causes the balance to become unstable or stiffen up along the travel if the stiffness at the weighing position, which is in the middle of the coil travel, is adjusted down to $0$.\\

(c) The stiffness needs to be constant with time, because changes over time result in a change in the restoring force of the mechanism, which leads to a drifting equilibrium position of the balance and therefore drift in the force signal, which might be limiting for the balance performance.
\\

At present, we are confident that a mechanism stiffness of $\leq\SI{0.01}{\newton \per \meter}$ will satisfy the above requirements. Experimental validation will follow in the near future. Optionally, we include a stiffness adjustment unit. This can be realized by applying a negative stiffness to the mechanism by means of, e.g., an inverted pendulum, astatic stiffness adjustment, or a negative spring restoring torque~\cite{Darnieder2018, Darnieder2019, Pratt2002}. 

\subsubsection*{Guiding of the coil}\label{subsection:GuidingCoil}

For QEMMS, the maximum horizontal parasitic motion along the vertical coil travel is \SI{10}{\micro \meter}. 
Furthermore, we are seeking to create a trajectory with the QEMMS mechanism that is repeatable between weighing and velocity mode as well as between each sweep in the velocity mode.

The coil speed in the velocity mode is measured with a heterodyne interferometer reflecting a laser beam off of a retroreflector that is attached to the moving coil (see figure~\ref{fig:KibbOperationVel}). The most basic criterion to run a measurement in the velocity mode is that the interference detector does not lose the beam signal in the sweeping phase. The laser beam of the interferometer needs to be fully reflected. Horizontal displacements and imperfections of the trajectory from the vertical can cause the beam to be clipped. However, before clipping the beam, other impacts to the measurement occur due to imperfections in the coil trajectory from the vertical~\cite{Sanchez2014, Haddad2016}. Limiting the explanations to the $x$-direction, these are proportional to the term $v_\mathrm{x}/v_\mathrm{z}$, where $v_\mathrm{x}$ is a horizontal and $v_\mathrm{z}$ is the vertical coil velocity. In appendix A, we discuss three biases that occur with horizontal motion. For all, the figure of merit is $v_\mathrm{x}/v_\mathrm{z}=\Delta x/\Delta z\approx \SI{1.67e-4}{}$. 

\subsubsection*{Low hysteretic force}\label{subsection:Hysteresis}

Non-linearities from mechanical hysteresis can be a problem in Kibble balances. If hysteresis is too large, then a precise measurement cannot be done because the this is a limitation in the resolution of the balance. 

The magnitude of a hysteretic force in general, depends on the load to the pivots and on excitations of the balance, e.g., due to a response of the balance during a mass transfer or the mechanism movement during the velocity mode. There are two general ways to reduce hysteretic forces in balances: (1) Minimizing the excursion during mass placement by adjusting mass lift speed and by optimizing feedback control. (2) Using an erasing procedure where the center pivot point is exercised in a damped sinusoidal motion after each excursion of the balance. However, a small part of hysteresis remains in the system. The goal is to have a remaining hysteretic loss in the mechanism equivalent to $\leq \SI{0.1}{\micro \gram}$.

The cause and magnitude of a hysteretic force is also driven by the type of mechanism used in the balance. All balance mechanisms can be divided based on their type of center pivot into two groups: knife edge-based and flexure pivot-based balances. In the QEMMS mechanism, flexure pivots will be employed.

The main advantage of a flexure over a knife edge with regards to hysteresis is that in the flexure only elastic deformation contributes to this effect, whereas in a knife edge elastic and plastic deformation need to be considered. After a change in the load and a deflection, a flexure shows anelastic behavior because it takes time to relax back to its internal equilibrium state. However, despite the similarities to the knife edge, the hysteretic forces in a flexure are known to be smaller. 
Reviewing literature, T.~Quinn~\cite{Quinn1992} published results for a \SI{1}{\kilo \gram} mass comparator stating that a flexure-based balance performs mass measurements with precision of up to six orders of magnitude better than a knife edge-based balance. Note that he admits that the knife edge-based balance was probably not ultimately optimized in terms of knife and flat materials and geometry.
~\cite{Quinn1992, Speake1987, Speake2018}

\begin{figure}[!htb]
    \centering
    \includegraphics[width=.55\columnwidth]{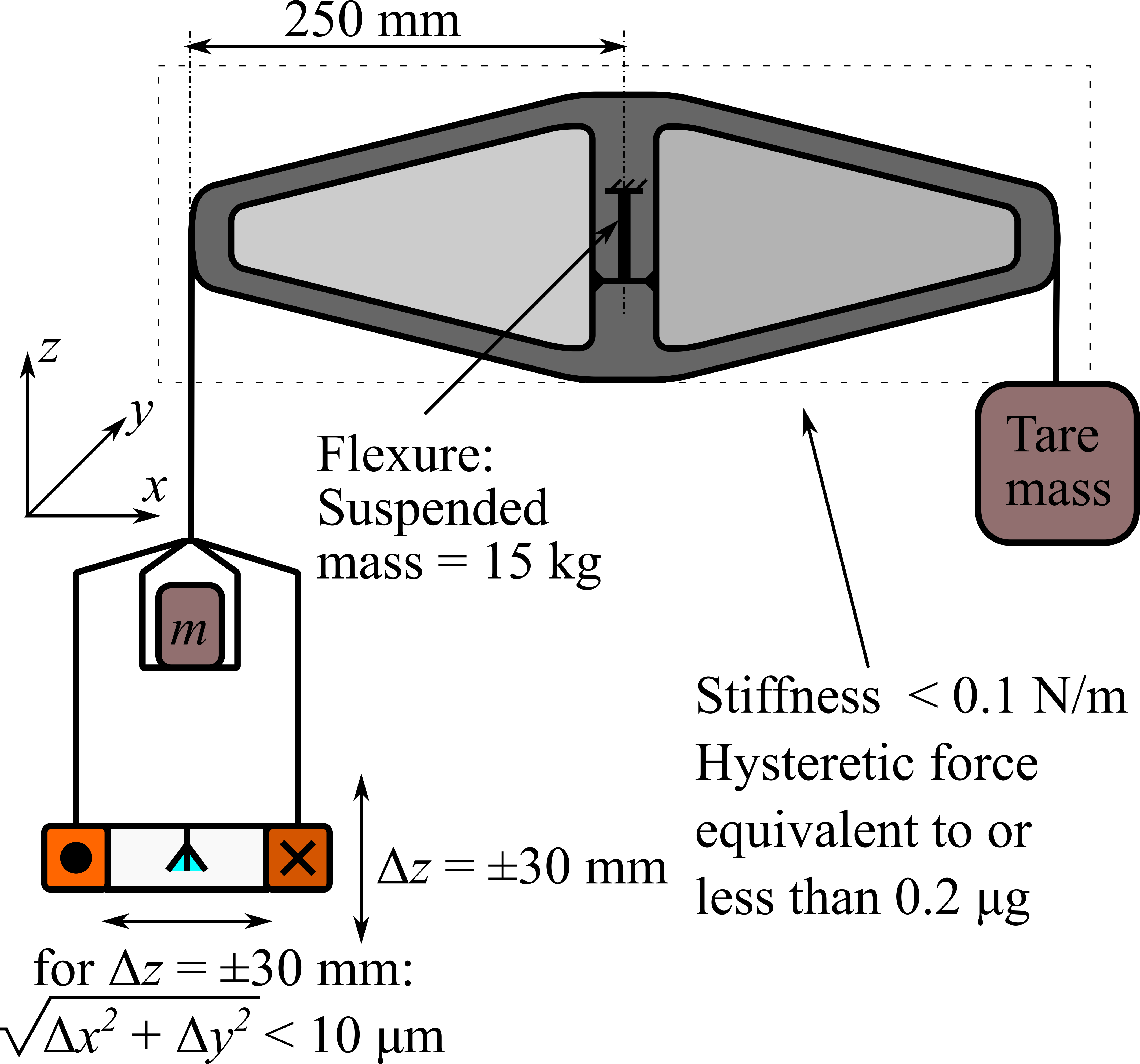}
    \caption{The requirements to the mechanism in the Kibble balance of the QEMMS summed up in a drawing.}
    \label{fig:KibbConceptDim}
\end{figure}

Note also that all published anelasticity data concerning flexures in precision balances were gathered using mass comparators which only use the weighing mode. The anelastic characteristic of flexures was never shown in Kibble balances where the weighing and the velocity mode are both performed by a single flexure mechanism. Conducting a series of experiments is necessary to learn about this effect because it cannot be quantitatively pre-determined during the design process. Unfortunately, theoretical calculation or simulation cannot provide quantitative knowledge regarding hysteretic forces because theories on this topic are not very well developed or validated. There are only general design recommendations that can be applied for minimizing the amount of anelasticity in flexures, for example: (1) designing the flexure as thin as possible in the region of bending, (2) designing it long and (3) using a flexure material with low internal damping coefficient. These recommendations minimize the geometric and elastic part of the stiffness, the internal material friction and therefore the energy stored and dissipated in the flexure during deflection, which is a measure for low hysteresis according to Speake~\cite{Speake2018}.

\subsubsection*{Carrying capacity and installation space} 

The central pivot in the QEMMS mechanism needs to support a total load of \SI{15}{\kilo \gram}. Furthermore, with this load it must be capable of rotating by $\pm \SI{7}{^\circ}$ to provide the desired $\pm \SI{30}{\milli \meter}$ of vertical travel, assuming the maximum dimension of a beam/wheel limited by the maximum installation space in QEMMS.
Since the Kibble balance in QEMMS is designed to be similar in size to commercial high precision vacuum mass comparators, we allow for a maximum horizontal dimension of the mechanism of $\approx \SI{500}{\milli \meter}$. This limits the maximum dimension of one beam or wheel arm in the balance to \SI{250}{\milli \meter} considering symmetric balance design. Also, the mechanism itself needs to be as compact as possible to keep free space for optical systems and a mass exchange unit in the vacuum chamber. The sensitivity of a balance scales as the square of the balance arm~\cite{Pratt2002}. Hence it should be as large as is practically allowed.

Figure~\ref{fig:KibbConceptDim} visualizes the above explained requirements in a drawing.

\subsection*{Design of the QEMMS mechanism}\label{section:NewMech}

\begin{figure}[!htb]
    \centering
    \includegraphics[width=.6\columnwidth]{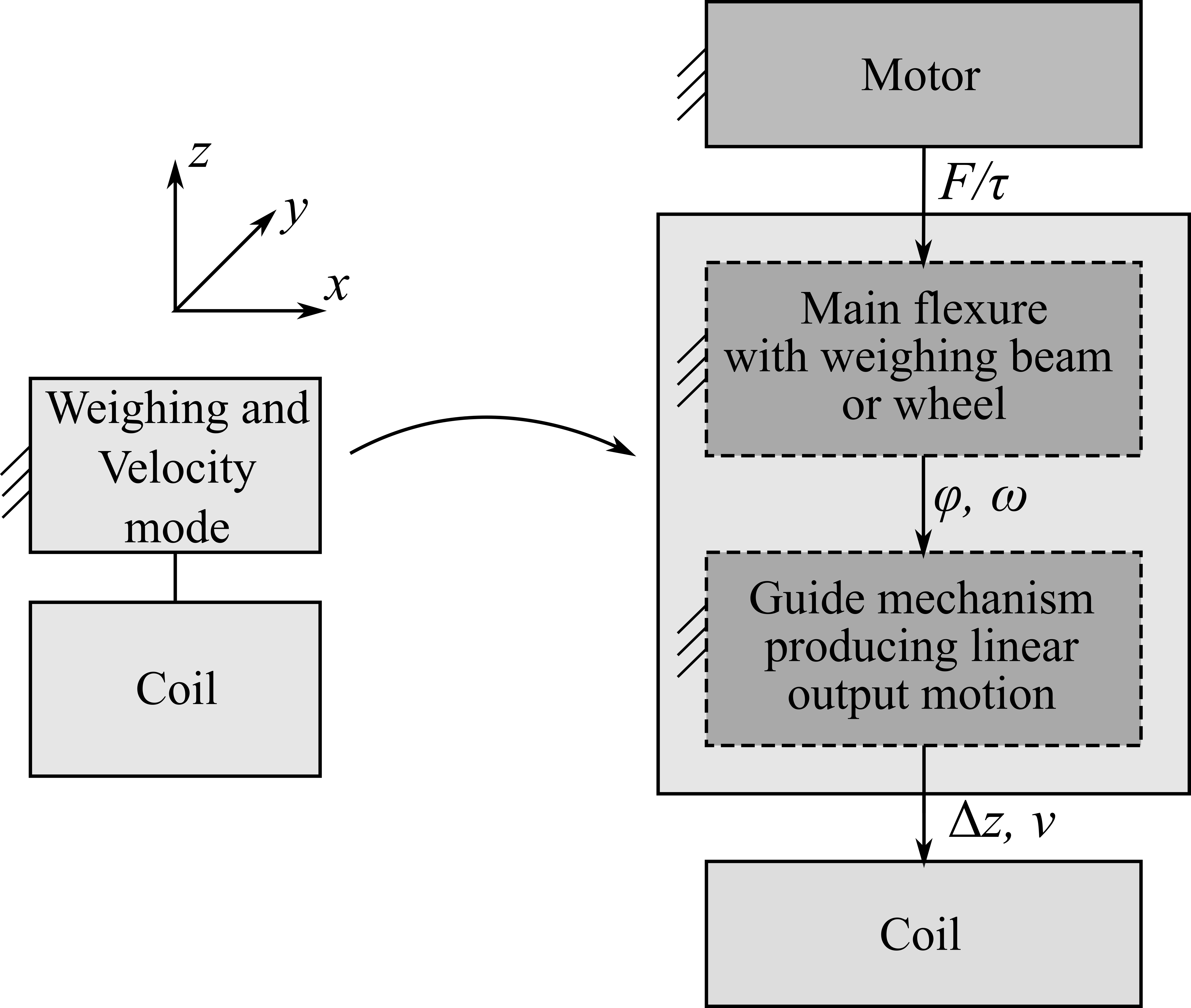}
    \caption{Close up of the weighing and velocity functionality. The motor is indicated to show where a driving force/torque for the velocity mode is put into the mechanism. Here, $F$ and $\tau$ denote a force or torque applied to the mechanism by the motor used to move the coil in the velocity mode. Furthermore, $\varphi$ and $\omega$ are the deflection angle and angular velocity of the main pivot. Finally, $\Delta z$ and $v$ are the translational displacement and velocity of the coil in vertical direction constrained by a guiding mechanism.}
    \label{fig:MechanismConcept}
\end{figure}

We are quite confident that a flexure-based mechanism design keeps the mechanical hysteresis in the QEMMS mechanism small and provides a repeatable movement. Furthermore, the new mechanism should be more compact, lighter, and require less auxiliary control features to define a one-dimensional coil trajectory than existing versions of Kibble balance mechanisms. Henceforth, we favor passive mechanical components built from flexure elements for defining the balance trajectory over auxiliary electrical components. Figure~\ref{fig:MechanismConcept} shows the separation of the QEMMS mechanism function in two sub-components integrated in a single mechanical system. One component is for balancing, the other one for defining a linear trajectory of the coil.

We will take a look into how to realize each functional component in figure~\ref{fig:MechanismConcept} with a flexure design and integrate them into a mechanism that satisfies the requirements for the QEMMS.

\subsubsection*{Main flexure} 

\begin{figure*}[!tbp]
    \begin{subfigure}[b]{0.42\columnwidth}
        \includegraphics[width=\columnwidth]{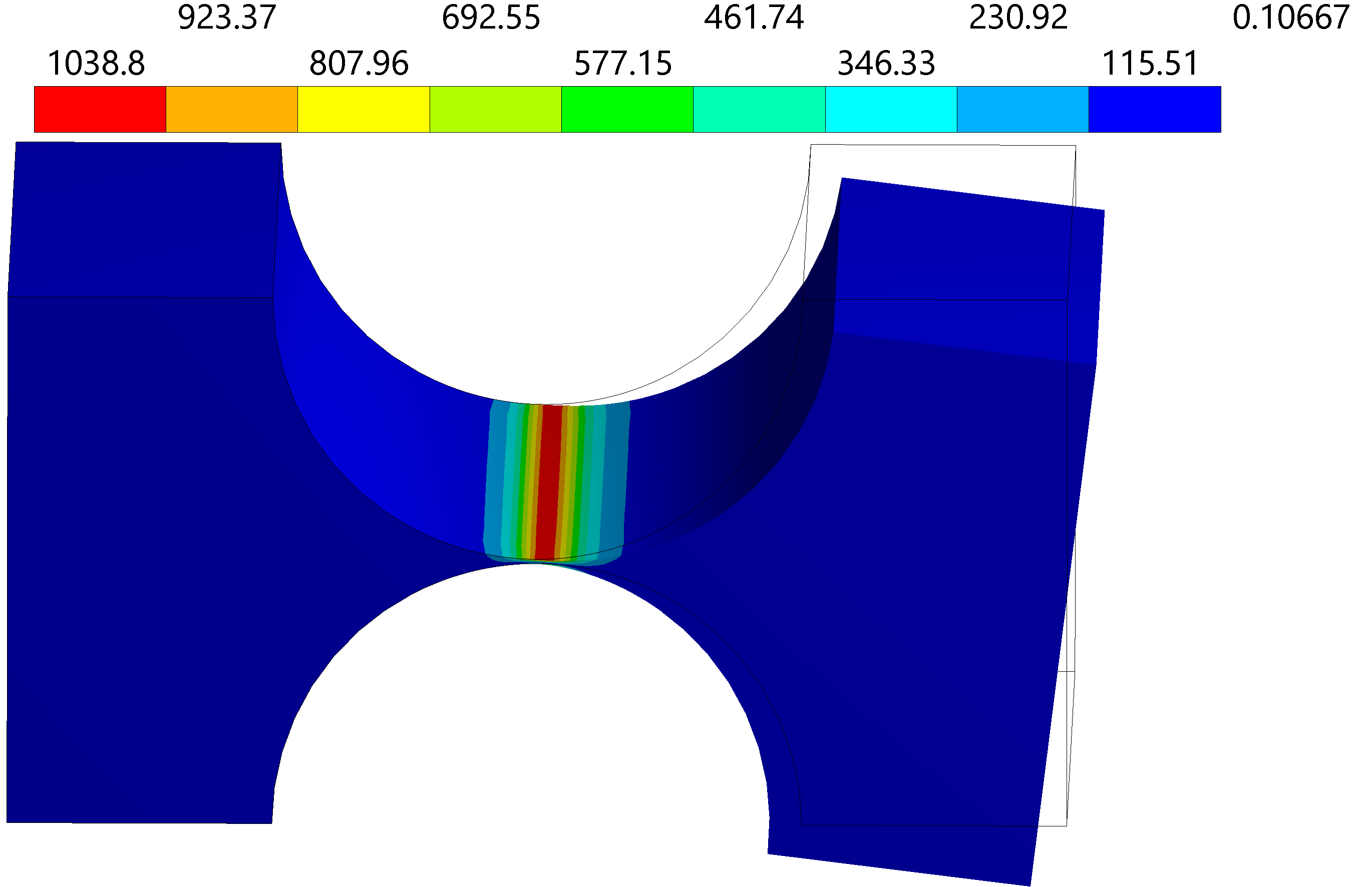}
    \end{subfigure}
    \hfill
    \begin{subfigure}[b]{0.42\columnwidth}
        \includegraphics[width=\columnwidth]{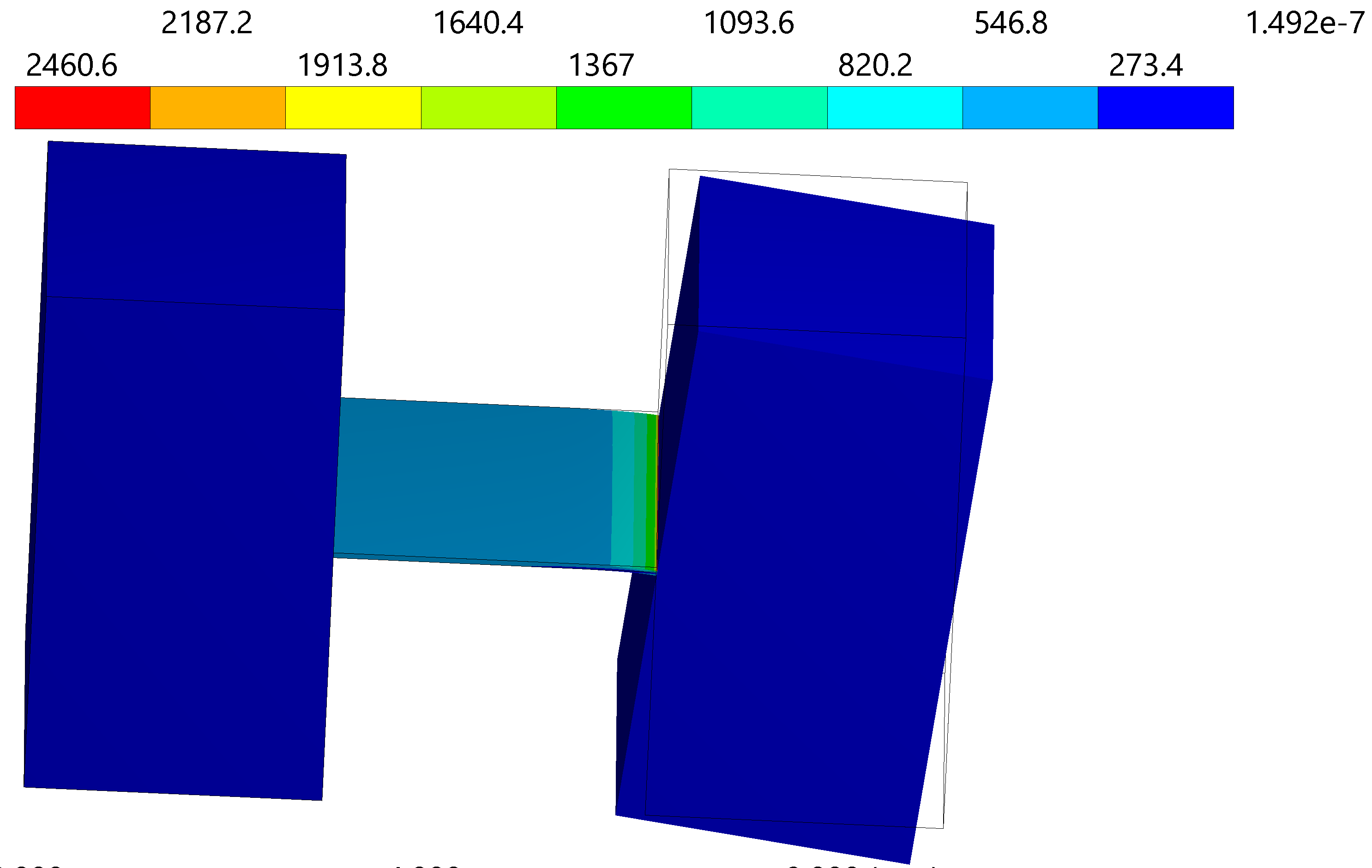}
    \end{subfigure}
    \caption{Finite element simulation of a conventional notch flexure with a notch radius of \SI{3}{\milli\meter} and a flat flexure with a length of \SI{3.25}{\milli\meter}. Both have a width of \SI{10}{\milli\meter} and a minimal notch thickness of \SI{0.05}{\milli\meter}. Boundary conditions: Each flexure is fixed supported at the left end and a rotation of \SI{7}{^\circ} is induced at the nodes at the right end. A point mass of \SI{15}{\kilogram}, which mimics the weight of the balance components, is attached on the right end. The color bars show the results for the maximum equivalent stress in the geometry after deflection in \SI{}{\mega \pascal}. In the flat flexure the high stress concentration is close to the right edge of the flat region.}
    \label{fig:FEA}
\end{figure*}

The trade-off in the design of the main flexure is to keep the maximum stress in the flexure at the maximum deflection angle at an acceptable level. A safety factor of $2$ to $3$ is desired. Also, a low elastic flexure stiffness is recommended in order to reduce the anelastic effect in the flexure after deflection~\cite{Speake2018}. This calls for a flexure as thin as possible.

Furthermore, a material with low internal damping/anelasticity must be used in order to avoid hysteretic effects upfront as much as possible. A metal with very high yield strength ($\approx \SI{1000}{MPa}$) and small anelastic after effect is a hardened Copper Beryllium alloy as used for flexures in balance experiments, e.g., in~\cite{Quinn1992}.

These materials are typically machined using wire electrical discharge machining, grinding, high speed milling or etching. In notch flexures, minimal flexure thicknesses of $\approx \SI{50}{\micro \meter}$ can be achieved. 

The main flexure needs to be carefully designed and analyzed by means of finite element analysis in order to evaluate stress-reducing parameters in the flexure accurately. One key factor to evaluate is the stress concentration near axis of rotation of the flexure due to the high axial load and high deflection. Widely used hinge geometries such as flat or conventional circular hinges have limitations for application in the QEMMS because they show high stress values derived from the simulations in figure~\ref{fig:FEA}. Maximum equivalent stresses of $>\SI{1000}{MPa}$ exceed the static yield strength for the high strength flexure material we chose ($\approx \SI{1000}{MPa}$). 

\begin{figure}[h!]
    \centering
    \includegraphics[width=.55\columnwidth]{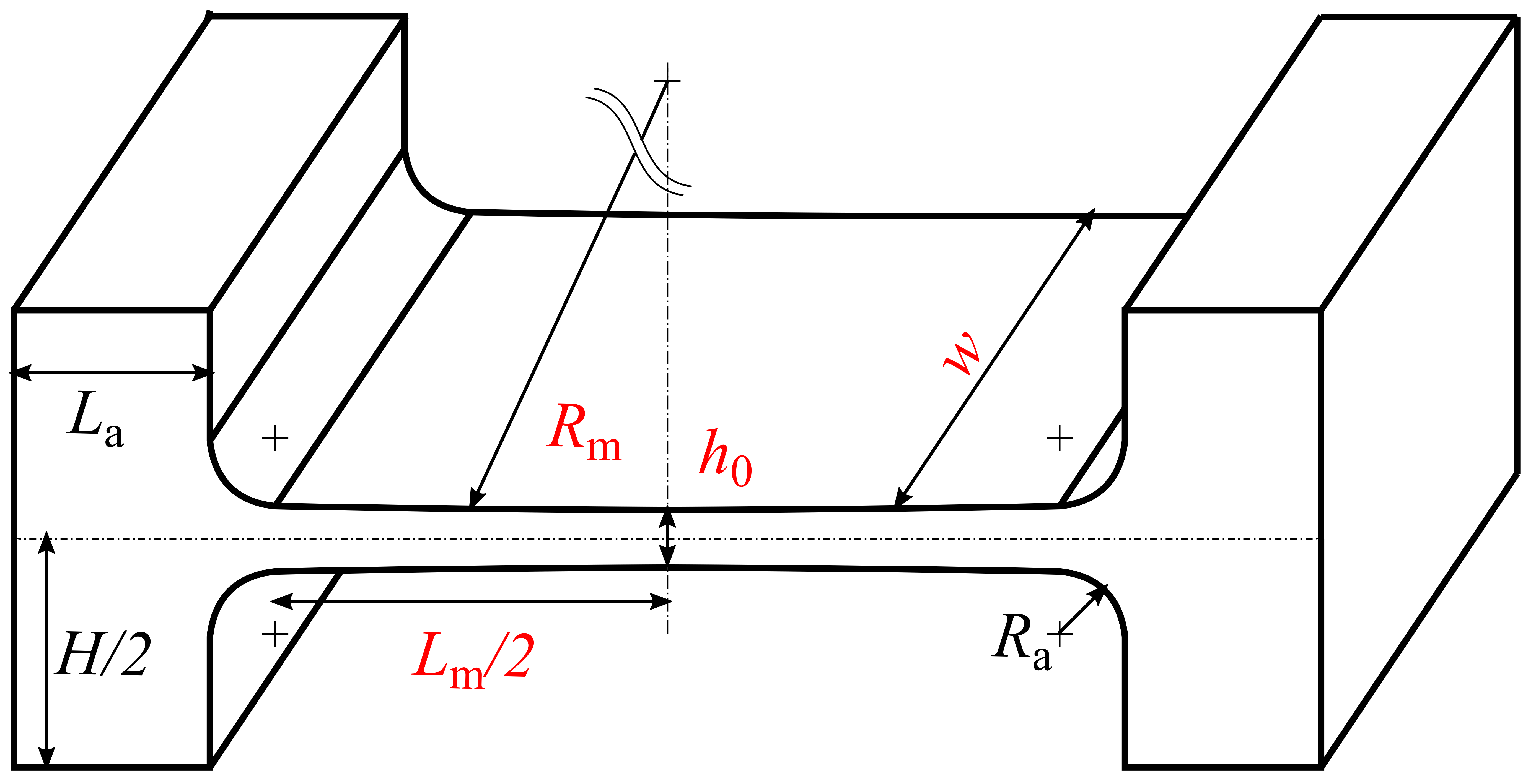}
    \caption{Visualization of the modified flexure geometry. As opposed to a regular circular flexure, the region of bending here has a very large radius $R_\mathrm{m}$ over the flexing length $L_\mathrm{m}$, with the characteristics $R_\mathrm{m} >> L_\mathrm{m}$. It has a minimal notch height $h_\mathrm{0}$ and a width $w$. $R_\mathrm{a}$ smoothens the transition from the bending region to the stiff part of the notch flexure. }
    \label{fig:flex_modified}
\end{figure}

In the investigated conventional flexure geometries, the bending radius is very small, which causes a high stress concentration in the region of flexing. Thus, we explore a modified flexure geometry that causes the bending radius of the flexure to become larger. This shows a positive effect to the maximum equivalent stress in the flexure. A representation for the modified geometry with an acceptable maximum stress value is shown in figure~\ref{fig:flex_modified}.

\begin{figure}[h!]
    \centering
    \includegraphics[width=.65\columnwidth]{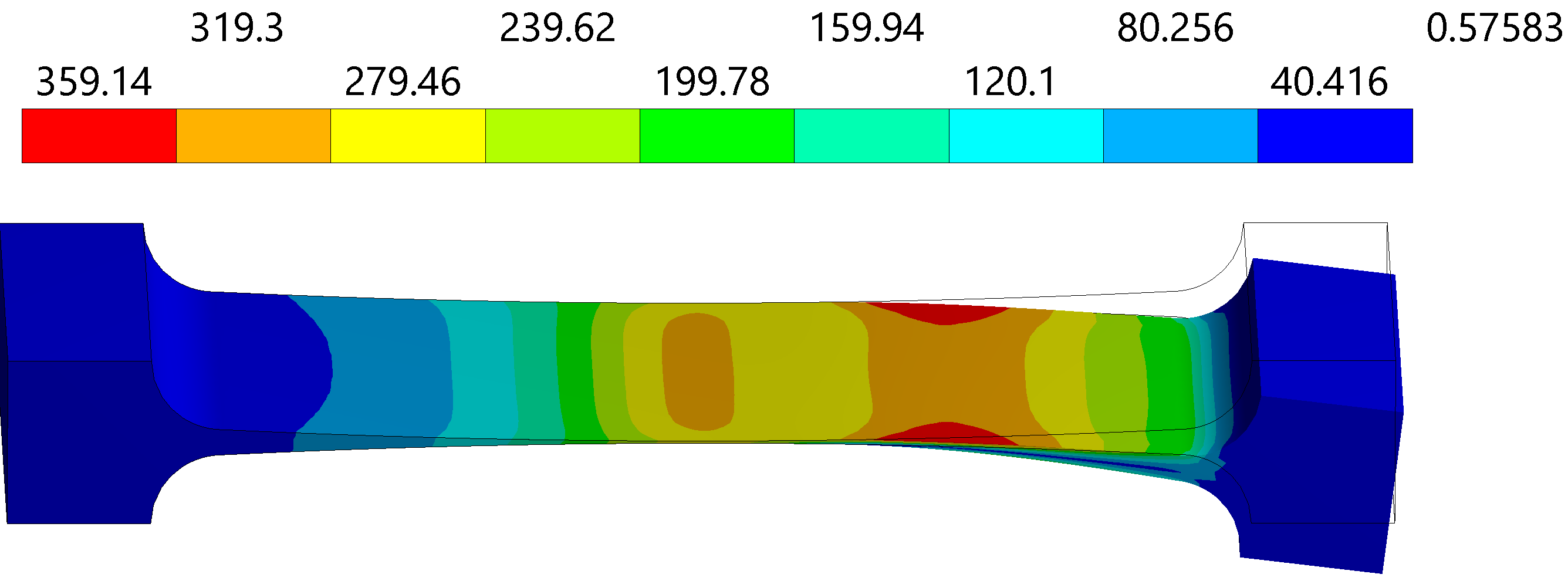}
    \caption{Finite element analysis stress plot of a representation of the modified flexure geometry. The parameters are $R_\mathrm{m}=\SI{200}{\milli\meter}$, $L_\mathrm{m}=\SI{20}{\milli\meter}$, $R_\mathrm{a}=\SI{1.5}{\milli\meter}$, $L_\mathrm{a}=\SI{3}{\milli\meter}$, $H=\SI{5}{\milli\meter}$, $w=\SI{10}{\milli\meter}$, and $h_\mathrm{0}=\SI{0.05}{\milli\meter}$. Boundary conditions are the same as in figure~\ref{fig:FEA}. The color bars show the results for the maximum equivalent stress in the flexure after deflection in \SI{}{\mega \pascal}. The simulation shows a maximum equivalent stress of $\SI{359}{\mega \pascal}$ in the flexure, which is about \SI{36}{\%} of the yield strength ($\approx \SI{1000}{MPa}$) of the flexure material.}
    \label{fig:new_flexure_stress}
\end{figure}

Figure~\ref{fig:new_flexure_stress} only represents one set of parameters for a flexure suitable for the application in the QEMMS mechanism. Other variations are possible, but further elaboration on this matter is not within the scope of this manuscript. The optimization of the geometry not only regarding the stress in the flexure but also the elastic stiffness is work in progress at NIST and will be published in a later on.

\subsubsection*{Kinematics}

In addition to the main pivot, a sub-mechanism for guiding with a proper link to the main flexure completes the kinematics of the balance. Examples, where a guiding mechanism is combined and linked with a balance beam can be found in, e.g., commercial weighing cells or in the Planck balance at Physikalisch-Technische Bundesanstalt (PTB)~\cite{Rothleitner2018}. An approach for an integrated design of main flexure and guiding mechanism in a parallel four bar linkage was used at NIST for electrostatic force balance experiments~\cite{Pratt2002, Keck2021}. The mechanisms in~\cite{Pratt2002, Rothleitner2018, Keck2021} show a systematic arc motion caused by the use of a parallel four bar linkage for guiding and use only small travel ranges on the order of micro meters during operation. Thus, these concepts can inspire a new design, but may not be directly used in the QEMMS due to the requirement for purely vertical motion of the coil over a relatively large travel. Other options for building flexure-based guiding mechanisms providing a linear motion need to be explored.

\begin{figure*}[h!]
    \centering
    \includegraphics[width=.9\columnwidth]{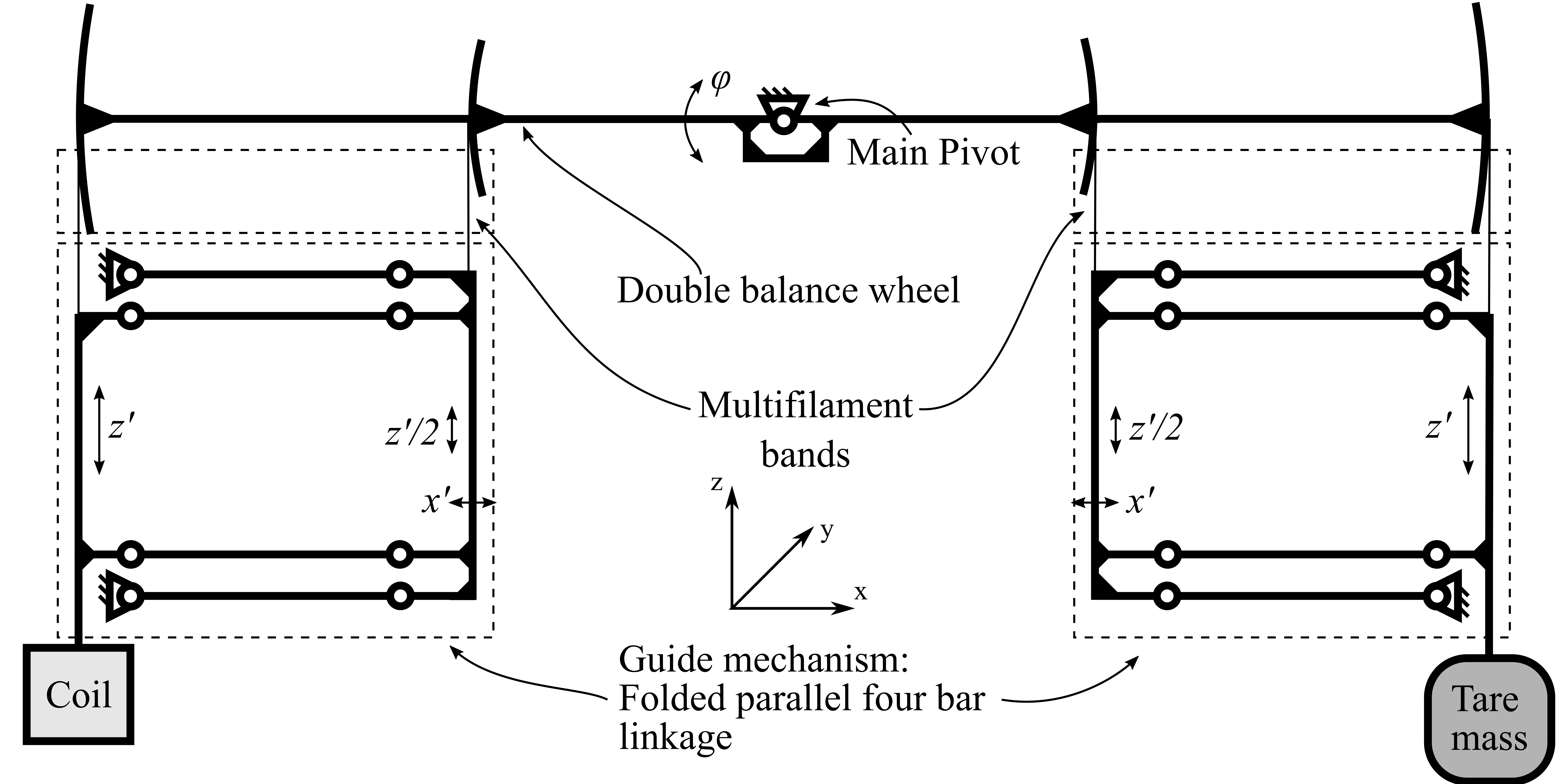}
    \caption{Conceptual visualization of the kinematics for the QEMMS mechanism. The principle of operation is based on a symmetric folded parallelogram linkage with external linkage. The drawing shows a representation of a wheel balance where bands roll off two wheels at the balancing structure.
    }
    \label{fig:QEMMS_mechanism_concept}
\end{figure*}

Since a single flexure hinge is always a rotating pivot, only a chain of multiple hinges can provide a (nearly) linear motion of a moving member. An overview and further references to useful literature about principles to combine a series of rotational pivots are provided in~\cite{Linss2015}. The dissertation thesis addresses the correction of unwanted parasitic motions as well as the change of the degree of freedom in a kinematic structure in order to achieve a certain trajectory. Such principles are, e.g., the design of parallel or serial kinematic structures, or applying nested or external linkage structures to avoid under-constraint in a mechanism~\cite{Panas2015}. An example based on a planar parallelogram linkage is provided in figure~\ref{fig:KinematicCompensation} in appendix B.

However, due to simplicity, correction of lateral error motions, compactness, and the convenience in machining of a planar flexure mechanism, we favor the folded parallelogram linkage as a guiding mechanism for the QEMMS. We will integrate a weighing beam/wheel as the external linkage as shown in option (4.2) in figure~\ref{fig:KinematicCompensation} (appendix B) and use this directly as a link to the main flexure. A visualization of the kinematic system for the QEMMS mechanism is shown in figure~\ref{fig:QEMMS_mechanism_concept}.

In a next step, a comparison of using a beam balance versus a wheel balance is addressed.

\subsubsection*{Beam balance vs. wheel balance}

There is a fundamental difference between the kinematics of the two balance types. The end points of the beam balance perform a rotational motion along a fixed circle. Only by the use of a further guiding mechanism, the movement of suspended components can take place parallel to the vector of gravity as shown in figure~\ref{fig:QEMMS_Simpl_Model} (appendix C). 

In contrast, with the wheel balance, a rolling of a band on the wheel is performed. This results, in theory, in a direct conversion of a rotational movement of the wheel into a linear displacement of suspended components parallel to the vector of the gravitational force. In order to achieve a further improvement in the quality of the linear motion and to avoid unwanted oscillations, a guiding mechanism as shown in figure~\ref{fig:QEMMS_mechanism_concept} can be used. The resulting overconstrained design can be dealt with by adjustment.

At first glance, this difference in use of a beam to a wheel balance seems to be negligible for application in a balance. Indeed, in a mass comparator, where the mechanism operates closely around its defined zero position, there is no theoretical functional advantage to either solution. All parasitic effects in the kinematics are small due to small deflections of the flexures. Convenience in machining and assembling of small planar (sometimes monolithic) mechanisms seems to be causing the favor for beam balances here. 

However, with the large travel required in the QEMMS, not all of the usually negligible effects in the mechanism remain harmless to the mechanical properties of the instrument. In fact, there is one impact that causes a major disadvantage of a beam balance compared to a wheel balance from a kinetic perspective: there is a horizontal force acting upon the connecting links between the guiding mechanism and the beam when the effective horizontal length of the balance beam shortens by the cosine of the deflection angle. 

This results in a non-linear stiffness term and parasitic forces to the guiding mechanism which cause unwanted deformations and error motions in the guide. An analysis of this effect is provided appendix C.\\

We could think of compensating the parasitic horizontal force on the stages by applying a symmetric compensation design approach by adding more components to a beam balance-based mechanism. However, this would increase the complexity of the mechanism to a level that we do not desire. The fact that the wheel balance prevents this horizontal force by design causes a favor of the wheel over the beam balance for the QEMMS mechanism.

\subsubsection*{Modular design}

\begin{figure}[h!]
    \centering
    \includegraphics[width=.95\columnwidth]{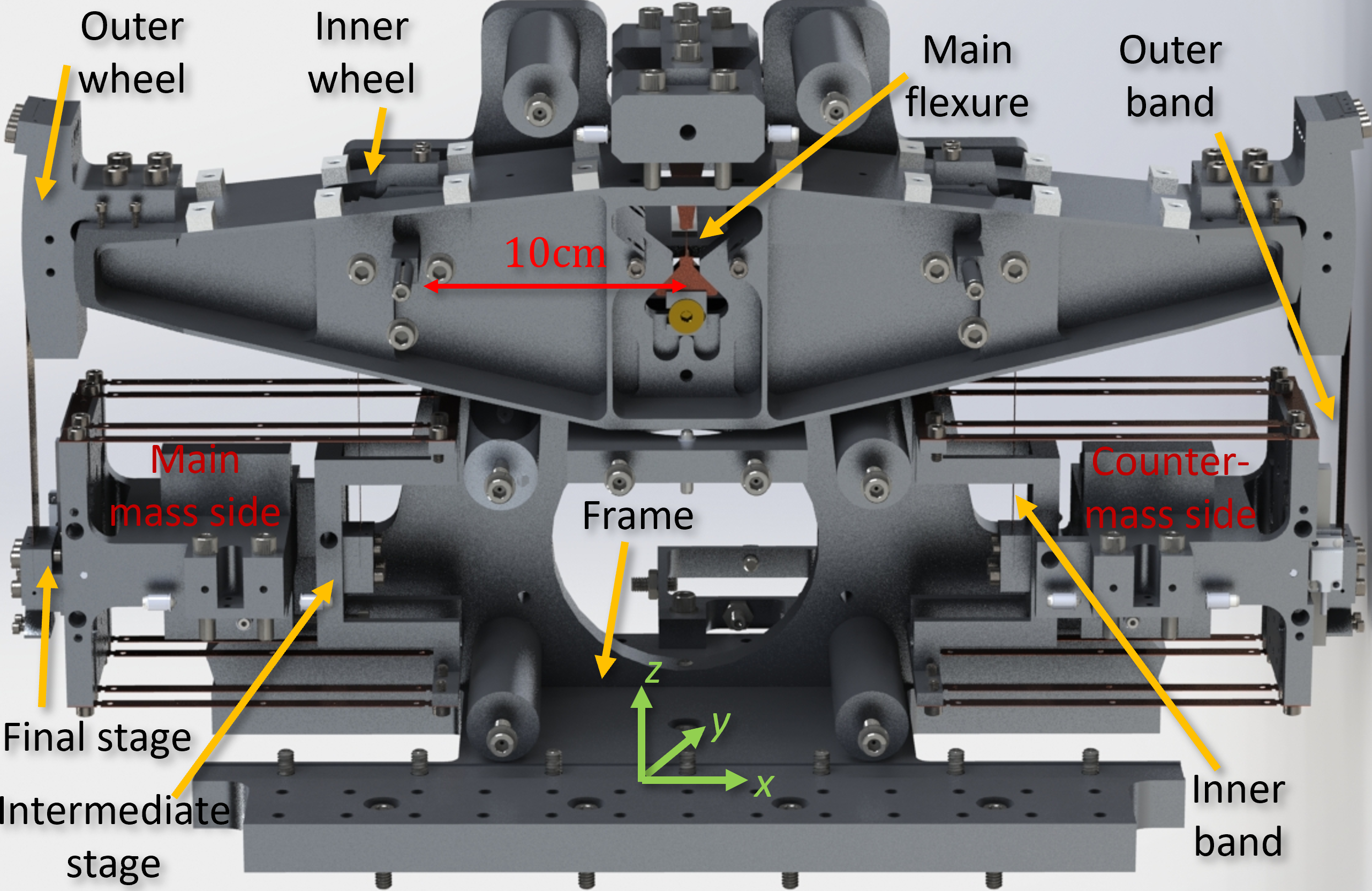}
    \caption{Render of the moveable parts of the QEMMS mechanism.}
    \label{fig:QEMMS_Mech_Render}
\end{figure}

The mechanism has been designed based on a wheel balance and a rendering of the frame and the moving parts can be seen in figure~\ref{fig:QEMMS_Mech_Render}. We favor a prototype of modular build to allow for variation of single components in experiments. Interesting elements for variation would be, e.g., the main flexure or the guiding mechanism flexures with view on the hysteretic properties of the mechanism or the guiding quality of the compliant structure.

The wheels are designed with wheel parts that are positioned and bolted down on a carrier beam structure. Multifilament bands are clamped, roll off the wheel surfaces and connect the intermediate and final stages of the guiding mechanisms with the inner and outer wheel. 

The modular design furthermore provides us with the opportunity to methodically investigate the effect of certain misalignments to the movement and hysteretic properties of the QEMMS mechanism. This helps to further clarify which design parameters matter not just from a theoretical, but from a practical point of view and highlights practical requirements to precision in assembling.

\section*{Conclusion}\label{Conclusion}

In this article we discuss the requirements and design for a new mechanism for the Kibble balance in the Quantum Electro-Mechanical Metrology Suite (QEMMS) currently under construction at NIST. Summarizing the explanations on the previous pages, we designed a fully mechanical and flexure-based balance mechanism that can execute both modes of operation in the Kibble balance experiment -- the weighing and the velocity mode. This mechanism constrains the trajectory of the moving components to a purely linear, one-dimensional motion by design without requiring auxiliary active control features. 

Experiments regarding the guiding quality and repeatability of the mechanism will show the technological limitations of the flexure design. Due to the modular design, we can optimize the mechanism after evaluation of experimental results methodically piece by piece to converge to a design with the properties we desire for the QEMMS mechanism.

A prudent design is to theoretically minimize the known factors contributing to the hysteretic effect, such as the spring constant and loss factor. However, ambiguity remains about the contribution of mechanical hysteresis to the uncertainty budget.

\newpage

\section*{Appendix A}\label{AppendixA}

Here, we discuss three errors introduced to the measurement with the Kibble balance in the velocity mode stemming from imperfections in the motion of the coil from the vertical. We limit the explanations to the $x$-direction and linear assumptions for simplicity. The errors describe systematic biases, that we are estimating with an upper limit value. In reality, once we build the experiment, we will measure all the imperfection terms in situ, e.g., the wave front distortion, the beam diameter, the horizontal motion of the balance mechanism, \ldots, and apply corrections for these biases.

\subsubsection*{Voltage bias}

Horizontal velocities also cause a bias $e_\mathrm{V}$ in the readouts of the induced voltage in the velocity mode. This stems from both horizontal forces $F_\mathrm{x}$ to the coil produced in the weighing mode -- when the electrical center of the coil is not aligned with the magnetic center of the permanent magnet -- and horizontal velocities in the velocity mode. The relative bias is expressed as follows

\begin{equation}
    e_\mathrm{V} =  \frac{F_\mathrm{x}}{F_\mathrm{z}} \frac{\Delta x}{\Delta z},
\end{equation}

where $F_\mathrm{z}$ is the force applied in the vertical direction by the magnet-coil system in weighing mode. The factor $F_\mathrm{x}/F_\mathrm{z}$ is typically on the order of \SI{1e-5}{}~\cite{Sanchez2014, Haddad2016}, thus this bias yields $e_\mathrm{V}=\SI{1.7e-9}{}$. However, this effect can be cancelled according to~\cite{Kibble2014} when employing a balance mechanism that performs the exact same motion in the weighing and the velocity mode, which is the goal for the QEMMS mechanism.

\subsubsection*{Beam shear bias}

A contribution to a bias in the velocity mode is the beam shear error $e_\mathrm{BS}$ in the interferometer. It occurs when the coil moves horizontal and the back-reflected interferometer beam gets displaced horizontally such that there is change in overlap between the reference beam and the beam reflected off of the moving retroreflector at the coil. The important values here are the wavelength of the laser, $\lambda$, the wave distortion at the optics -- we assume $\lambda/10$ -- the beam shear/horizontal coil motion, $\Delta x$, and the diameter of the laser beam in the interferometer, $d_\mathrm{Beam}$. The equation for this relative bias derived from~\cite{Niebauer1995} is

\begin{equation}
    e_\mathrm{BS} = \frac{\lambda}{10} \frac{2}{d_\mathrm{Beam}} \frac{\Delta x}{\Delta z}.
\end{equation}\label{eq:beamshear}

The wavelength of the laser is $\lambda = \SI{633}{\nano\meter}$ with a beam diameter of $d_\mathrm{Beam} = \SI{6}{\milli\meter}$ which yields a bias of $e_\mathrm{BS} = \SI{3.52e-9}{}$.

\subsubsection*{Velocity bias}

Another measurement bias, $e_\mathrm{v}$, comes from horizontal velocities during the velocity mode. These are derived from a misalignment of the interferometer with respect to gravity, $\alpha$, and from the horizontal motion, $\Delta x$, along the vertical coil travel $\Delta z$~\cite{Haddad2016, Gillespie1997}.
The equation for the bias is

\begin{equation}
    e_\mathrm{v} = \alpha \frac{\Delta x}{\Delta z}.
\end{equation}

Assuming a reasonable value $\alpha = \SI{45}{\micro \radian}$, we get a relative measurement bias of $e_\mathrm{v} = \SI{7.5e-9}{}$, which is on the order of magnitude we desire to have.

\section*{Appendix B}\label{AppendixB}

\begin{figure*}[t]
    \centering
    \includegraphics[width=.85\columnwidth]{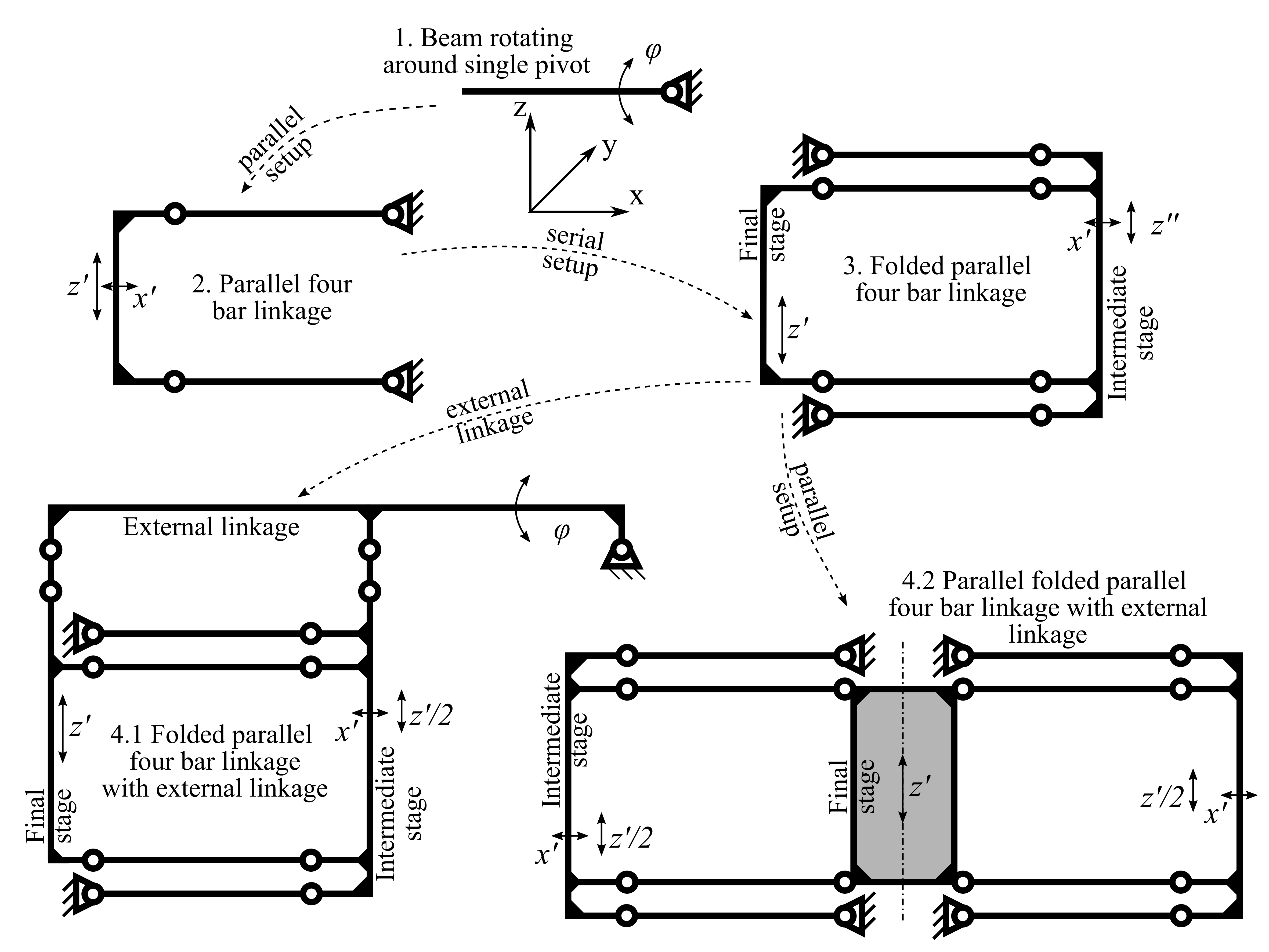}
    \caption{In first approximation, flexure mechanisms can be modelled with rigid links connected to rotational pivots. This diagram shows an example for the correction of unwanted motions in the kinematic structure of mechanisms consisting of rotational joints. Starting with a purely rotational joint (1.) and ending with a linear motion provided by a chain of rotational joints, (4.1) and (4.2). The principles of serial and parallel setup of pivots as well as an external linkage are applied.}
    \label{fig:KinematicCompensation}
\end{figure*}

Figure~\ref{fig:KinematicCompensation} explains design principles to correct for motion imperfections in kinematic chains transitioning from a purely rotational to a purely translational motion of a moving member in a mechanism.

A purely rotational motion (1.) is turned into a quasi-linear motion of a plane with a systematic horizontal part ($x'$) through parallel setup of two beams as in a parallel four bar linkage (2.). Now we can use a serial (folded) chain of these linkages. By moving ($z''$) half the distance of ($z'$) we can correct for the parasitic horizontal part of the motion in the final stage. However, this is a movable system with two degrees of freedom, where the movement of the intermediate stage on the right ($z''$) is not coupled to the final stage on the left ($z'$) (3.). Furthermore, if the equation $z'' = 2z'$ is not fulfilled here, there is going to be a certain $x'$ in the final trajectory. In a final step, applying an external linkage (4.1) or parallel setup (4.2) we can constrain the motion to a single degree of freedom, as long as the condition $z'' = 2z'$ is fulfilled by design. 

Option (4.1) was used, e.g., as the mechanism in an electrostatic balance at NIST~\cite{Kramar2002}, and option (4.2) as the dedicated guiding mechanism for the Mark II Kibble balance of the Federal Institute of Metrology (METAS) of Switzerland~\cite{Cosandier2014}.

\section*{Appendix C}\label{AppendixC}

\begin{figure}[h!]
    \centering
    \includegraphics[width=.55\columnwidth]{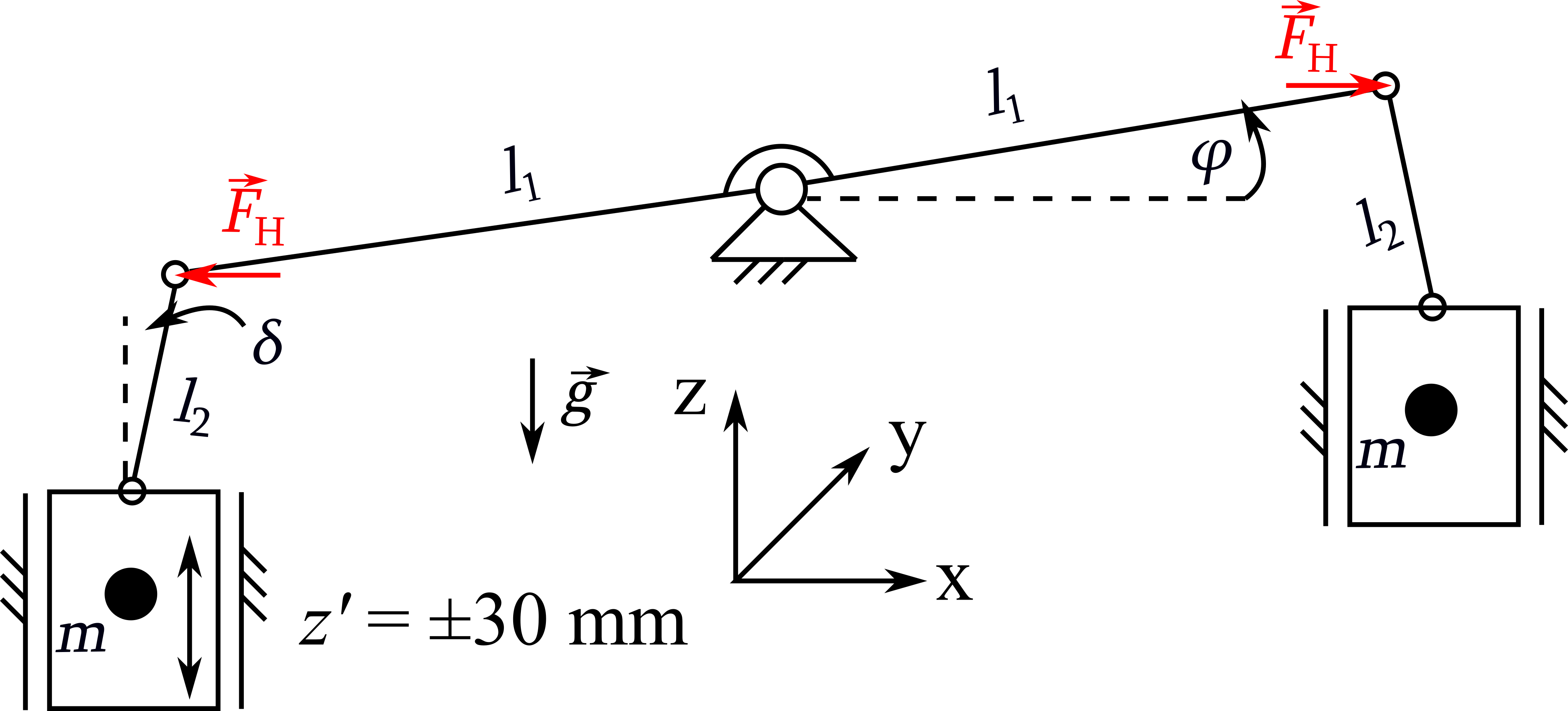}
    \caption{Simplified model of a beam balance combined with guiding mechanisms on both sides.}
    \label{fig:QEMMS_Simpl_Model}
\end{figure}

We derive the non-linear stiffness from the simplified model of a beam balance connected with a guiding mechanism on both sides in figure~\ref{fig:QEMMS_Simpl_Model}.

The horizontal force $F_\mathrm{H}$ depending on the rotational deflection $\varphi$ is described with

\begin{equation}
    F_\mathrm{H} (\varphi) = m \, g \, \frac{l_\mathrm{1}}{l_\mathrm{2}} \, \frac{\varphi^2}{2},
\end{equation}

truncating the Taylor approximations for sine and cosine after the second order term.

The lever arm $z' = l_\mathrm{1} \, \varphi$ introduces a moment $M_\mathrm{H} (\varphi)$ through each side of the beam according to 

\begin{equation}
    M_\mathrm{H} (\varphi) = m \, g \, \frac{l_\mathrm{1}^2}{l_\mathrm{2}} \, \varphi^3,
\end{equation}

which points out a third order dependency of the induced moment on the deflection angle $\varphi$.

For simplification, we assume that we compensate a constant, elastic part of the flexure stiffness in the mechanism entirely by using, e.g., an inverted pendulum. Building the derivative of the previous equation due to $\varphi$ yields the non-linear rotational stiffness term induced by the parasitic horizontal force $F_\mathrm{H}$

\begin{equation}
    K = \frac{\mbox{d}M}{\mbox{d}\varphi} = 3 \, m \, g \, \frac{l_\mathrm{1}^2}{l_\mathrm{2}} \, \varphi^2.\label{eq:Nonlin_Stiff}
\end{equation}

\begin{figure}[h!]
    \centering
    \includegraphics[width=.7\columnwidth]{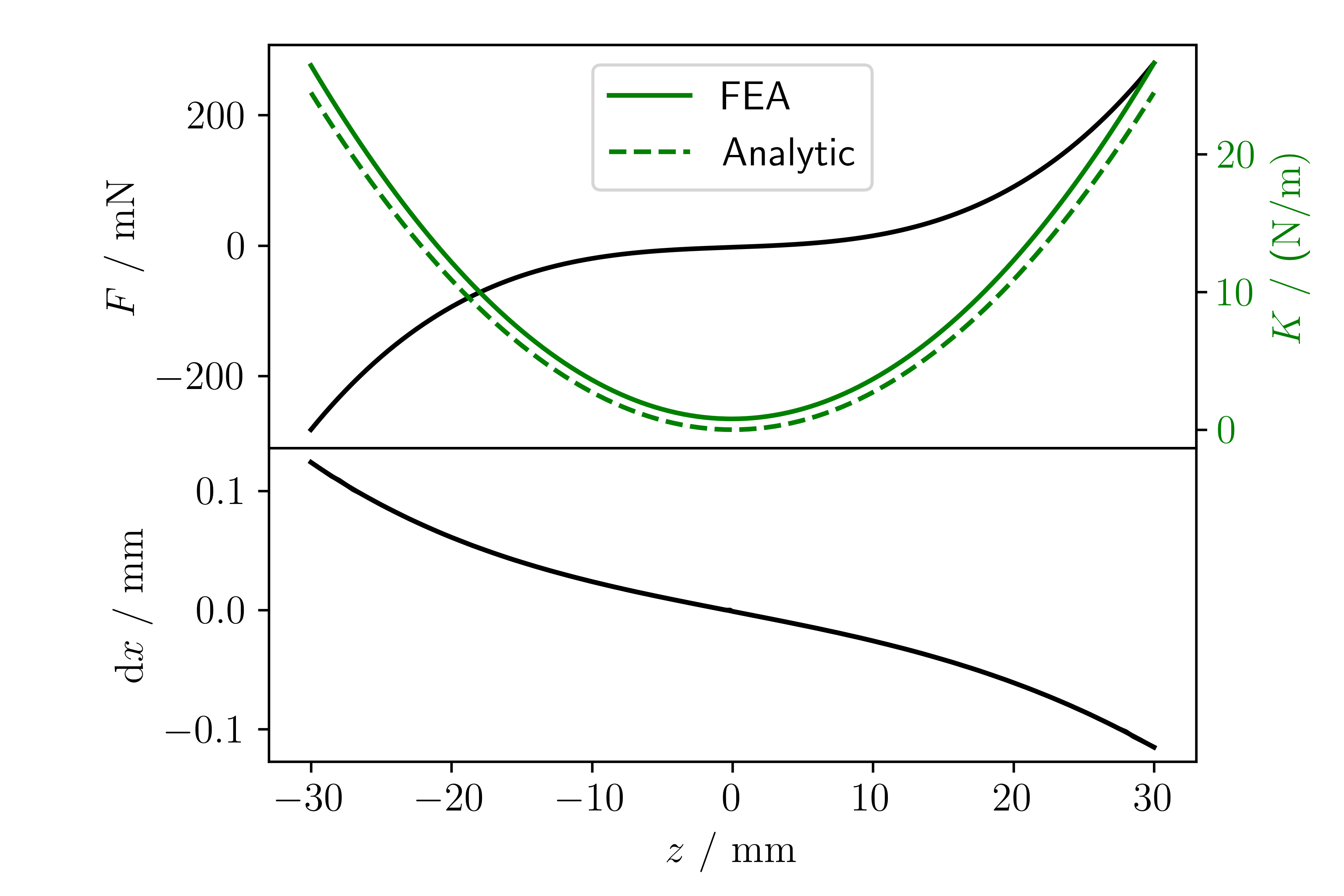}
    \caption{Results of a finite element analysis (FEA) at a beam balance model of the QEMMS mechanism. Upper plot: the left axis shows a third order fit of the displacement force and the right axis the derived stiffness along the vertical ($z$) displacement axis. The solid lines are from finite element analysis and the dashed line is from the analytical model for the mechanism stiffness. The parameters for the simulation are: $m = \SI{6}{\kilogram}$, $g = \SI{9.81}{\newton \per \kilogram}$, $l_\mathrm{1} = \SI{250}{\milli \meter}$, $l_\mathrm{2} = \SI{100}{\milli \meter}$. Lower plot: shows the horizontal trajectory error of the final stage of the mechanism $\mbox{d}x$ over $z$.}
    \label{fig:QEMMS_Sim_Stiff}
\end{figure}

\begin{figure*}[h!]
    \centering
    \includegraphics[width=.95\columnwidth]{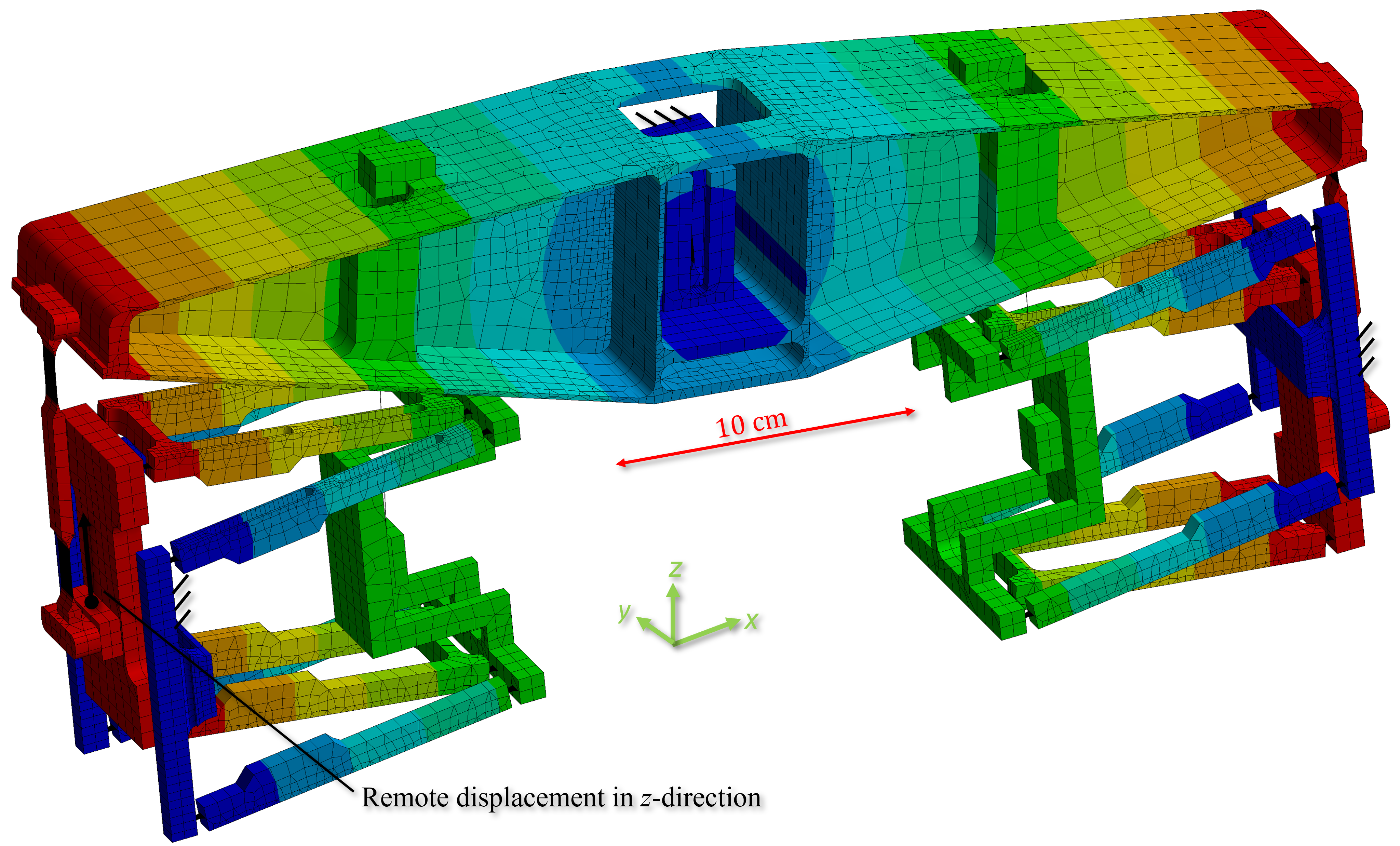}
    \caption{Total deformation plot of a finite element analysis at the beam balance model for the QEMMS mechanism. The final stage was displaced step-wise $\pm \SI{30}{\milli \meter}$ in $z$-direction and the force reaction as well as the error motion in $x$-direction was monitored. The result is shown in figure~\ref{fig:QEMMS_Sim_Stiff}.}
    \label{fig:FEA_Beam}
\end{figure*}

Unfortunately, this term becomes highly dominant with larger deflections as can be seen in the upper plot in figure~\ref{fig:QEMMS_Sim_Stiff}. With reasonable values for the system parameters shown in figure~\ref{fig:QEMMS_Simpl_Model}, a study at a simplified beam balance-based mechanism model using finite element analysis was conducted. Figure~\ref{fig:FEA_Beam} shows the model and boundary conditions therefore. 

The force over displacement curve of the final stage of the mechanism was monitored and compared to the analytical model. The rotational stiffness in equation~\ref{eq:Nonlin_Stiff} can be transferred to a linear stiffness at the guiding mechanism in vertical direction by dividing equation~\ref{eq:Nonlin_Stiff} by the square of the balance lever arm $l_\mathrm{1}^2$. Furthermore, $\varphi \approx \Delta z / l_\mathrm{1}$.

Note that the constant offset of the dashed from the solid green line in figure~\ref{fig:QEMMS_Sim_Stiff} stems from a residual constant stiffness part in the simulation model, which can, in practice, be compensated by adjusting the center of mass of the beam~\cite{Pratt2002}.

Furthermore, the horizontal force $F_\mathrm{H}$ acting upon the flexure-based guiding mechanism affects the quality of guiding. The simulation yielding the results for the stiffness of the mechanism in figure~\ref{fig:QEMMS_Sim_Stiff} also provides insights to this horizontal trajectory error, $\mbox{d}x$, along the vertical displacement axis in $z$-direction displayed in the lower plot in figure~\ref{fig:QEMMS_Sim_Stiff}. We see that the total horizontal error motion is approximately $20$ times larger than allowed.

\newpage

%%%%%%%%%%%%%%%%%%%%%%%%%%%%%%%%%%%%%%%%%%%%%%
%%                                          %%
%% Backmatter begins here                   %%
%%                                          %%
%%%%%%%%%%%%%%%%%%%%%%%%%%%%%%%%%%%%%%%%%%%%%%

\begin{backmatter}

\end{backmatter}


\begin{thebibliography}{6}
\ifx \bisbn   \undefined \def \bisbn  #1{ISBN #1}\fi
\ifx \binits  \undefined \def \binits#1{#1}\fi
\ifx \bauthor  \undefined \def \bauthor#1{#1}\fi
\ifx \batitle  \undefined \def \batitle#1{#1}\fi
\ifx \bjtitle  \undefined \def \bjtitle#1{#1}\fi
\ifx \bvolume  \undefined \def \bvolume#1{\textbf{#1}}\fi
\ifx \byear  \undefined \def \byear#1{#1}\fi
\ifx \bissue  \undefined \def \bissue#1{#1}\fi
\ifx \bfpage  \undefined \def \bfpage#1{#1}\fi
\ifx \blpage  \undefined \def \blpage #1{#1}\fi
\ifx \burl  \undefined \def \burl#1{\textsf{#1}}\fi
\ifx \doiurl  \undefined \def \doiurl#1{\textsf{#1}}\fi
\ifx \betal  \undefined \def \betal{\textit{et al.}}\fi
\ifx \binstitute  \undefined \def \binstitute#1{#1}\fi
\ifx \binstitutionaled  \undefined \def \binstitutionaled#1{#1}\fi
\ifx \bctitle  \undefined \def \bctitle#1{#1}\fi
\ifx \beditor  \undefined \def \beditor#1{#1}\fi
\ifx \bpublisher  \undefined \def \bpublisher#1{#1}\fi
\ifx \bbtitle  \undefined \def \bbtitle#1{#1}\fi
\ifx \bedition  \undefined \def \bedition#1{#1}\fi
\ifx \bseriesno  \undefined \def \bseriesno#1{#1}\fi
\ifx \blocation  \undefined \def \blocation#1{#1}\fi
\ifx \bsertitle  \undefined \def \bsertitle#1{#1}\fi
\ifx \bsnm \undefined \def \bsnm#1{#1}\fi
\ifx \bsuffix \undefined \def \bsuffix#1{#1}\fi
\ifx \bparticle \undefined \def \bparticle#1{#1}\fi
\ifx \barticle \undefined \def \barticle#1{#1}\fi
\ifx \bconfdate \undefined \def \bconfdate #1{#1}\fi
\ifx \botherref \undefined \def \botherref #1{#1}\fi
\ifx \url \undefined \def \url#1{\textsf{#1}}\fi
\ifx \bchapter \undefined \def \bchapter#1{#1}\fi
\ifx \bbook \undefined \def \bbook#1{#1}\fi
\ifx \bcomment \undefined \def \bcomment#1{#1}\fi
\ifx \oauthor \undefined \def \oauthor#1{#1}\fi
\ifx \citeauthoryear \undefined \def \citeauthoryear#1{#1}\fi
\ifx \endbibitem  \undefined \def \endbibitem {}\fi
\ifx \bconflocation  \undefined \def \bconflocation#1{#1}\fi
\ifx \arxivurl  \undefined \def \arxivurl#1{\textsf{#1}}\fi
\csname PreBibitemsHook\endcsname

%%% 1
\bibitem{koon}
\begin{barticle}
\bauthor{\bsnm{Koonin}, \binits{E.V.}},
\bauthor{\bsnm{Altschul}, \binits{S.F.}},
\bauthor{\bsnm{Bork}, \binits{P.}}:
\batitle{Brca1 protein products: functional motifs}.
\bjtitle{Nat. Genet.}
\bvolume{13},
\bfpage{266}--\blpage{267}
(\byear{1996})
\end{barticle}
\endbibitem

%%% 2
\bibitem{xjon}
\begin{bchapter}
\bauthor{\bsnm{Jones}, \binits{X.}}:
\bctitle{Zeolites and synthetic mechanisms}.
In: \beditor{\bsnm{Smith}, \binits{Y.}} (ed.)
\bbtitle{Proceedings of the First National Conference on Porous Sieves: 27-30
  June 1996; Baltimore},
pp. \bfpage{16}--\blpage{27}
(\byear{1996})
\end{bchapter}
\endbibitem

%%% 3
\bibitem{marg}
\begin{bbook}
\bauthor{\bsnm{Margulis}, \binits{L.}}:
\bbtitle{Origin of Eukaryotic Cells}.
\bpublisher{Yale University Press},
\blocation{New Haven}
(\byear{1970})
\end{bbook}
\endbibitem

%%% 4
\bibitem{schn}
\begin{bchapter}
\bauthor{\bsnm{Schnepf}, \binits{E.}}:
\bctitle{From prey via endosymbiont to plastids: comparative studies in
  dinoflagellates}.
In: \beditor{\bsnm{Lewin}, \binits{R.A.}} (ed.)
\bbtitle{Origins of Plastids},
\bedition{2nd} edn.,
pp. \bfpage{53}--\blpage{76}.
\bpublisher{Chapman and Hall},
\blocation{New York}
(\byear{1993})
\end{bchapter}
\endbibitem

%%% 5
\bibitem{koha}
\begin{botherref}
\oauthor{\bsnm{Kohavi}, \binits{R.}}:
Wrappers for performance enhancement and obvious decision graphs.
PhD thesis,
Stanford University, Computer Science Department
(1995)
\end{botherref}
\endbibitem

%%% 6
\bibitem{issnic}
\begin{botherref}
\oauthor{\bsnm{{ISSN International Centre}}}:
The ISSN register
(2006).
\url{http://www.issn.org}
Accessed Accessed 20 Feb 2007
\end{botherref}
\endbibitem

\end{thebibliography}

\newcommand{\BMCxmlcomment}[1]{}

\BMCxmlcomment{

<refgrp>

<bibl id="B1">
  <title><p>BRCA1 protein products: functional motifs</p></title>
  <aug>
    <au><snm>Koonin</snm><fnm>E V</fnm></au>
    <au><snm>Altschul</snm><fnm>S F</fnm></au>
    <au><snm>Bork</snm><fnm>P</fnm></au>
  </aug>
  <source>Nat. Genet.</source>
  <pubdate>1996</pubdate>
  <volume>13</volume>
  <fpage>266</fpage>
  <lpage>267</lpage>
</bibl>

<bibl id="B2">
  <title><p>Zeolites and synthetic mechanisms</p></title>
  <aug>
    <au><snm>Jones</snm><fnm>X</fnm></au>
  </aug>
  <source>Proceedings of the First National Conference on Porous Sieves: 27-30
  June 1996; Baltimore</source>
  <editor>Y Smith</editor>
  <pubdate>1996</pubdate>
  <fpage>16</fpage>
  <lpage>27</lpage>
</bibl>

<bibl id="B3">
  <title><p>Origin of Eukaryotic Cells</p></title>
  <aug>
    <au><snm>Margulis</snm><fnm>L</fnm></au>
  </aug>
  <publisher>New Haven: Yale University Press</publisher>
  <pubdate>1970</pubdate>
</bibl>

<bibl id="B4">
  <title><p>From prey via endosymbiont to plastids: comparative studies in
  dinoflagellates</p></title>
  <aug>
    <au><snm>Schnepf</snm><fnm>E</fnm></au>
  </aug>
  <source>Origins of Plastids</source>
  <publisher>New York: Chapman and Hall</publisher>
  <editor>R A Lewin</editor>
  <edition>2</edition>
  <pubdate>1993</pubdate>
  <fpage>53</fpage>
  <lpage>76</lpage>
</bibl>

<bibl id="B5">
  <title><p>Wrappers for performance enhancement and obvious decision
  graphs</p></title>
  <aug>
    <au><snm>Kohavi</snm><fnm>R</fnm></au>
  </aug>
  <source>PhD thesis</source>
  <publisher>Stanford University, Computer Science Department</publisher>
  <pubdate>1995</pubdate>
</bibl>

<bibl id="B6">
  <title><p>The ISSN register</p></title>
  <aug>
    <au><cnm>{ISSN International Centre}</cnm></au>
  </aug>
  <pubdate>2006</pubdate>
  <url>http://www.issn.org</url>
</bibl>

</refgrp>
} % end of \BMCxmlcomment


\begin{thebibliography}{30}
\ifx \bisbn   \undefined \def \bisbn  #1{ISBN #1}\fi
\ifx \binits  \undefined \def \binits#1{#1}\fi
\ifx \bauthor  \undefined \def \bauthor#1{#1}\fi
\ifx \batitle  \undefined \def \batitle#1{#1}\fi
\ifx \bjtitle  \undefined \def \bjtitle#1{#1}\fi
\ifx \bvolume  \undefined \def \bvolume#1{\textbf{#1}}\fi
\ifx \byear  \undefined \def \byear#1{#1}\fi
\ifx \bissue  \undefined \def \bissue#1{#1}\fi
\ifx \bfpage  \undefined \def \bfpage#1{#1}\fi
\ifx \blpage  \undefined \def \blpage #1{#1}\fi
\ifx \burl  \undefined \def \burl#1{\textsf{#1}}\fi
\ifx \doiurl  \undefined \def \doiurl#1{\textsf{#1}}\fi
\ifx \betal  \undefined \def \betal{\textit{et al.}}\fi
\ifx \binstitute  \undefined \def \binstitute#1{#1}\fi
\ifx \binstitutionaled  \undefined \def \binstitutionaled#1{#1}\fi
\ifx \bctitle  \undefined \def \bctitle#1{#1}\fi
\ifx \beditor  \undefined \def \beditor#1{#1}\fi
\ifx \bpublisher  \undefined \def \bpublisher#1{#1}\fi
\ifx \bbtitle  \undefined \def \bbtitle#1{#1}\fi
\ifx \bedition  \undefined \def \bedition#1{#1}\fi
\ifx \bseriesno  \undefined \def \bseriesno#1{#1}\fi
\ifx \blocation  \undefined \def \blocation#1{#1}\fi
\ifx \bsertitle  \undefined \def \bsertitle#1{#1}\fi
\ifx \bsnm \undefined \def \bsnm#1{#1}\fi
\ifx \bsuffix \undefined \def \bsuffix#1{#1}\fi
\ifx \bparticle \undefined \def \bparticle#1{#1}\fi
\ifx \barticle \undefined \def \barticle#1{#1}\fi
\ifx \bconfdate \undefined \def \bconfdate #1{#1}\fi
\ifx \botherref \undefined \def \botherref #1{#1}\fi
\ifx \url \undefined \def \url#1{\textsf{#1}}\fi
\ifx \bchapter \undefined \def \bchapter#1{#1}\fi
\ifx \bbook \undefined \def \bbook#1{#1}\fi
\ifx \bcomment \undefined \def \bcomment#1{#1}\fi
\ifx \oauthor \undefined \def \oauthor#1{#1}\fi
\ifx \citeauthoryear \undefined \def \citeauthoryear#1{#1}\fi
\ifx \endbibitem  \undefined \def \endbibitem {}\fi
\ifx \bconflocation  \undefined \def \bconflocation#1{#1}\fi
\ifx \arxivurl  \undefined \def \arxivurl#1{\textsf{#1}}\fi
\csname PreBibitemsHook\endcsname

%%% 1
\bibitem{Kibble1976}
\begin{bchapter}
\bauthor{\bsnm{Kibble}, \binits{B.}}:
\bctitle{A measurement of the gyromagnetic ratio of the proton by the strong
  field method}.
In: \bbtitle{Atomic Masses and Fundamental Constants},
vol. \bseriesno{5},
pp. \bfpage{545}--\blpage{551}
(\byear{1976})
\end{bchapter}
\endbibitem

%%% 2
\bibitem{BIPM2019}
\begin{botherref}
The {BIPM} and the metre convention: The {I}nternational {S}ystem of {U}nits
  {(SI)},
117--216
(2019)
\end{botherref}
\endbibitem

%%% 3
\bibitem{Tiesinga2021}
\begin{botherref}
\oauthor{\bsnm{Tiesinga}, \binits{E.}},
\oauthor{\bsnm{Mohr}, \binits{P.}},
\oauthor{\bsnm{Newell}, \binits{D.}},
\oauthor{\bsnm{Taylor}, \binits{B.}}:
Codata recommended values of the fundamental physical constants: 2018.
Rev. Mod. Phys.
\textbf{93}
(2021)
\end{botherref}
\endbibitem

%%% 4
\bibitem{Delahaye2003}
\begin{barticle}
\bauthor{\bsnm{Delahaye}, \binits{F.}},
\bauthor{\bsnm{Jeckelmann}, \binits{B.}}:
\batitle{Revised technical guidelines for reliable dc measurements of the
  quantized hall resistance}.
\bjtitle{Metrologia}
\bvolume{40},
\bfpage{217}--\blpage{240}
(\byear{2003})
\end{barticle}
\endbibitem

%%% 5
\bibitem{Jeckelmann2001}
\begin{barticle}
\bauthor{\bsnm{Jeckelmann}, \binits{B.}},
\bauthor{\bsnm{Jeanneret}, \binits{B.}}:
\batitle{The quantum hall effect as an electrical resistance standard}.
\bjtitle{Rep. Prog. Phys.}
\bvolume{64},
\bfpage{1603}--\blpage{1658}
(\byear{2001})
\end{barticle}
\endbibitem

%%% 6
\bibitem{Hamilton2000}
\begin{barticle}
\bauthor{\bsnm{Hamilton}, \binits{C.}}:
\batitle{Josephson voltage standards}.
\bjtitle{Rev. Sci. Instrum.}
\bvolume{71},
\bfpage{3611}--\blpage{3626}
(\byear{2000})
\end{barticle}
\endbibitem

%%% 7
\bibitem{Haddad2016b}
\begin{barticle}
\bauthor{\bsnm{Haddad}, \binits{D.}},
\bauthor{\bsnm{Seifert}, \binits{F.}},
\bauthor{\bsnm{Chao}, \binits{L.}},
\bauthor{\bsnm{Li}, \binits{S.}},
\bauthor{\bsnm{Newell}, \binits{D.}},
\bauthor{\bsnm{Pratt}, \binits{J.}},
\bauthor{\bsnm{Williams}, \binits{C.}},
\bauthor{\bsnm{Schlamminger}, \binits{S.}}:
\batitle{Bridging classical and quantum mechanics}.
\bjtitle{Metrologia}
\bvolume{53},
\bfpage{83}--\blpage{85}
(\byear{2016})
\end{barticle}
\endbibitem

%%% 8
\bibitem{Robinson2016a}
\begin{barticle}
\bauthor{\bsnm{Robinson}, \binits{I.}},
\bauthor{\bsnm{Schlamminger}, \binits{S.}}:
\batitle{The watt or {K}ibble balance: a technique for implementing the new
  {SI} definition of the unit of mass}.
\bjtitle{Metrologia}
\bvolume{53},
\bfpage{46}--\blpage{74}
(\byear{2016})
\end{barticle}
\endbibitem

%%% 9
\bibitem{Kruskopf2016}
\begin{botherref}
\oauthor{\bsnm{Kruskopf}, \binits{M.}},
\oauthor{\bsnm{Pakdehi}, \binits{K.} \bsuffix{DM.and~Pierz}},
\oauthor{\bsnm{Wundrack}, \binits{S.}},
\oauthor{\bsnm{Stosch}, \binits{R.}},
\oauthor{\bsnm{Dziomba}, \binits{T.}},
\oauthor{\bsnm{Götz}, \binits{M.}},
\oauthor{\bsnm{Baringhaus}, \binits{J.}},
\oauthor{\bsnm{Aprojanz}, \binits{J.}},
\oauthor{\bsnm{Tegenkamp}, \binits{C.}},
\oauthor{\bsnm{Lidzba}, \binits{J.}}:
Comeback of epitaxial graphene for electronics: large-area growth of
  bilayer-free graphene on sic. 2d materials.
2D Materials
\textbf{3}
(2016)
\end{botherref}
\endbibitem

%%% 10
\bibitem{Panna2021}
\begin{botherref}
\oauthor{\bsnm{Panna}, \binits{A.}},
\oauthor{\bsnm{Hu}, \binits{I.}},
\oauthor{\bsnm{Kruskopf}, \binits{M.}},
\oauthor{\bsnm{Patel}, \binits{D.}},
\oauthor{\bsnm{Jarrett}, \binits{D.}},
\oauthor{\bsnm{Liu}, \binits{C.}},
\oauthor{\bsnm{Payagala}, \binits{S.}},
\oauthor{\bsnm{Saha}, \binits{D.}},
\oauthor{\bsnm{Rigosi}, \binits{A.}},
\oauthor{\bsnm{Newell}, \binits{D.}},
\oauthor{\bsnm{Liang}, \binits{C.}},
\oauthor{\bsnm{Elmquist}, \binits{R.}}:
Graphene quantum {H}all effect parallel resistance arrays.
PHYSICAL REVIEW B
\textbf{103}(075408)
(2021)
\end{botherref}
\endbibitem

%%% 11
\bibitem{Rigosi2019}
\begin{barticle}
\bauthor{\bsnm{Rigosi}, \binits{A.}},
\bauthor{\bsnm{Panna}, \binits{A.}},
\bauthor{\bsnm{Payagala}, \binits{S.}},
\bauthor{\bsnm{Kruskopf}, \binits{M.}},
\bauthor{\bsnm{Kraft}, \binits{M.}},
\bauthor{\bsnm{Jones}, \binits{G.}},
\bauthor{\bsnm{Wu}, \binits{B.}},
\bauthor{\bsnm{Lee}, \binits{H.}},
\bauthor{\bsnm{Yang}, \binits{Y.}},
\bauthor{\bsnm{Hu}, \binits{J.}},
\bauthor{\bsnm{Jarrett}, \binits{D.}},
\bauthor{\bsnm{Newell}, \binits{D.}},
\bauthor{\bsnm{Elmquist}, \binits{R.}}:
\batitle{Graphene devices for tabletop and high-current quantized {H}all
  resistance standards}.
\bjtitle{IEEE Trans. Instrum. Meas.}
\bvolume{68},
\bfpage{1870}--\blpage{1878}
(\byear{2019})
\end{barticle}
\endbibitem

%%% 12
\bibitem{Marangoni2020}
\begin{barticle}
\bauthor{\bsnm{Marangoni}, \binits{R.}},
\bauthor{\bsnm{Haddad}, \binits{D.}},
\bauthor{\bsnm{Seifert}, \binits{F.}},
\bauthor{\bsnm{Chao}, \binits{L.}},
\bauthor{\bsnm{Newell}, \binits{D.}},
\bauthor{\bsnm{Schlamminger}, \binits{S.}}:
\batitle{Magnet system for the quantum electromechanical metrology suite}.
\bjtitle{IEEE Trans. Instrum. Meas.}
\bvolume{69},
\bfpage{5736}--\blpage{5744}
(\byear{2020})
\end{barticle}
\endbibitem

%%% 13
\bibitem{Marangoni2020b}
\begin{bchapter}
\bauthor{\bsnm{Marangoni}, \binits{R.}},
\bauthor{\bsnm{Haddad}, \binits{D.}},
\bauthor{\bsnm{Seifert}, \binits{F.}},
\bauthor{\bsnm{Chao}, \binits{L.}},
\bauthor{\bsnm{Pratt}, \binits{J.}},
\bauthor{\bsnm{Newell}, \binits{D.}},
\bauthor{\bsnm{Schlamminger}, \binits{S.}}:
\bctitle{Design of the {K}ibble balance for the {QEMMS}}.
In: \bbtitle{2020 Conference on Precision Electromagnetic Measurements (CPEM)},
pp. \bfpage{1}--\blpage{2}
(\byear{2020})
\end{bchapter}
\endbibitem

%%% 14
\bibitem{Darnieder2018}
\begin{barticle}
\bauthor{\bsnm{Darnieder}, \binits{M.}},
\bauthor{\bsnm{Pabst}, \binits{M.}},
\bauthor{\bsnm{Wenig}, \binits{R.}},
\bauthor{\bsnm{Zentner}, \binits{L.}},
\bauthor{\bsnm{Theska}, \binits{R.}},
\bauthor{\bsnm{Fr\"ohlich}, \binits{T.}}:
\batitle{Static behavior of weighing cells}.
\bjtitle{Journal of Sensors and Sensor Systems}
\bvolume{2},
\bfpage{587}--\blpage{600}
(\byear{2018})
\end{barticle}
\endbibitem

%%% 15
\bibitem{Darnieder2019}
\begin{bchapter}
\bauthor{\bsnm{Darnieder}, \binits{M.}},
\bauthor{\bsnm{Pabst}, \binits{M.}},
\bauthor{\bsnm{Fr\"ohlich}, \binits{T.}},
\bauthor{\bsnm{Zentner}, \binits{L.}},
\bauthor{\bsnm{Theska}, \binits{R.}}:
\bctitle{Mechanical properties of an adjustable weighing cell prototype}.
In: \bbtitle{Euspen’s 19th International Conference and Exhibition}
(\byear{2019})
\end{bchapter}
\endbibitem

%%% 16
\bibitem{Pratt2002}
\begin{bchapter}
\bauthor{\bsnm{Pratt}, \binits{J.}},
\bauthor{\bsnm{Newell}, \binits{D.}},
\bauthor{\bsnm{Kramar}, \binits{J.}}:
\bctitle{A flexure balance with adjustable restoring torque for nanonewton
  force measurement}.
In: \bbtitle{International Measurement Confederation}
(\byear{2002})
\end{bchapter}
\endbibitem

%%% 17
\bibitem{Sanchez2014}
\begin{barticle}
\bauthor{\bsnm{Sanchez}, \binits{C.}},
\bauthor{\bsnm{Wood}, \binits{B.}},
\bauthor{\bsnm{Green}, \binits{R.}},
\bauthor{\bsnm{Liard}, \binits{J.}},
\bauthor{\bsnm{Inglis}, \binits{D.}}:
\batitle{A determination of {P}lanck’s constant using the {NRC} watt
  balance}.
\bjtitle{Metrologia}
\bvolume{51},
\bfpage{5}--\blpage{14}
(\byear{2014})
\end{barticle}
\endbibitem

%%% 18
\bibitem{Haddad2016}
\begin{botherref}
\oauthor{\bsnm{Haddad}, \binits{D.}},
\oauthor{\bsnm{Seifert}, \binits{F.}},
\oauthor{\bsnm{Chao}, \binits{L.}},
\oauthor{\bsnm{Li}, \binits{S.}},
\oauthor{\bsnm{Newell}, \binits{D.}},
\oauthor{\bsnm{Pratt}, \binits{J.}},
\oauthor{\bsnm{Williams}, \binits{C.}},
\oauthor{\bsnm{Schlamminger}, \binits{S.}}:
Invited article: A precise instrument to determine the {P}lanck constant, and
  the future kilogram
\textbf{87}
(2016)
\end{botherref}
\endbibitem

%%% 19
\bibitem{Quinn1992}
\begin{barticle}
\bauthor{\bsnm{Quinn}, \binits{T.}}:
\batitle{The beam balance as an instrument for very precise weighing}.
\bjtitle{Meas. Sci. Technol.}
\bvolume{3},
\bfpage{141}--\blpage{159}
(\byear{1992})
\end{barticle}
\endbibitem

%%% 20
\bibitem{Speake1987}
\begin{barticle}
\bauthor{\bsnm{Speake}, \binits{C.}}:
\batitle{Fundamental limits to mass comparison by means of a beam balance}.
\bjtitle{Proceedings of the Royal Society of London. A. Mathematical and
  Physical Sciences}
\bvolume{414},
\bfpage{333}--\blpage{358}
(\byear{1987})
\end{barticle}
\endbibitem

%%% 21
\bibitem{Speake2018}
\begin{barticle}
\bauthor{\bsnm{Speake}, \binits{C.}}:
\batitle{Anelasticity in flexure strips revisited}.
\bjtitle{Metrologia}
\bvolume{55},
\bfpage{114}--\blpage{119}
(\byear{2018})
\end{barticle}
\endbibitem

%%% 22
\bibitem{Rothleitner2018}
\begin{barticle}
\bauthor{\bsnm{Rothleitner}, \binits{C.}},
\bauthor{\bsnm{Schleichert}, \binits{J.}},
\bauthor{\bsnm{Rogge}, \binits{N.}},
\bauthor{\bsnm{Günther}, \binits{L.}},
\bauthor{\bsnm{Vasilyan}, \binits{S.}},
\bauthor{\bsnm{Hilbrunner}, \binits{F.}},
\bauthor{\bsnm{Knopf}, \binits{D.}},
\bauthor{\bsnm{Fröhlich}, \binits{T.}},
\bauthor{\bsnm{Härtig}, \binits{F.}}:
\batitle{The {P}lanck-{B}alance{\textemdash}using a fixed value of the {P}lanck
  constant to calibrate {E1}/{E2}-weights}.
\bjtitle{Meas. Sci. Technol.}
\bvolume{29},
\bfpage{074003}
(\byear{2018})
\end{barticle}
\endbibitem

%%% 23
\bibitem{Keck2021}
\begin{barticle}
\bauthor{\bsnm{Keck}, \binits{L.}},
\bauthor{\bsnm{Shaw}, \binits{G.}},
\bauthor{\bsnm{Theska}, \binits{R.}},
\bauthor{\bsnm{Schlamminger}, \binits{S.}}:
\batitle{Design of an electrostatic balance mechanism to measure optical power
  of 100 k{W}}.
\bjtitle{IEEE Trans. Instrum. Meas.}
\bvolume{70},
\bfpage{1}--\blpage{9}
(\byear{2021})
\end{barticle}
\endbibitem

%%% 24
\bibitem{Linss2015}
\begin{botherref}
\oauthor{\bsnm{Lin{\ss}}, \binits{S.}}:
Ein {B}eitrag zur geometrischen {G}estaltung und {O}ptimierung prismatischer
  {F}estkörpergelenke in nachgiebigen {K}oppelmechanismen.
PhD thesis,
Technische Universi\"at Ilmenau, Fachgebiet Nachgiebige Systeme
(2015)
\end{botherref}
\endbibitem

%%% 25
\bibitem{Panas2015}
\begin{barticle}
\bauthor{\bsnm{Panas}, \binits{R.}},
\bauthor{\bsnm{Hopkins}, \binits{J.}}:
\batitle{Eliminating underconstraint in double parallelogram flexure
  mechanisms}.
\bjtitle{Journal of Mechanical Design}
\bvolume{137},
\bfpage{1}--\blpage{9}
(\byear{2015})
\end{barticle}
\endbibitem

%%% 26
\bibitem{Kibble2014}
\begin{barticle}
\bauthor{\bsnm{Kibble}, \binits{B.}},
\bauthor{\bsnm{Robinson}, \binits{I.}}:
\batitle{Principles of a new generation of simplified and accurate watt
  balances}.
\bjtitle{Metrologia}
\bvolume{51},
\bfpage{132}--\blpage{139}
(\byear{2014})
\end{barticle}
\endbibitem

%%% 27
\bibitem{Niebauer1995}
\begin{botherref}
\oauthor{\bsnm{Niebauer}, \binits{T.M.}},
\oauthor{\bsnm{Sasagawa}, \binits{G.S.}},
\oauthor{\bsnm{Fallel}, \binits{J.E.}},
\oauthor{\bsnm{Hilt}, \binits{R.}},
\oauthor{\bsnm{Klopping}, \binits{E.}}:
A new generation of absolute gravimeters.
Metrologia
\textbf{32}
(1995)
\end{botherref}
\endbibitem

%%% 28
\bibitem{Gillespie1997}
\begin{barticle}
\bauthor{\bsnm{Gillespie}, \binits{G.}},
\bauthor{\bsnm{Fujii}, \binits{K.}},
\bauthor{\bsnm{Newell}, \binits{D.B.}},
\bauthor{\bsnm{Olsen}, \binits{P.T.}},
\bauthor{\bsnm{Picard}, \binits{O.A.}},
\bauthor{\bsnm{Steiner}, \binits{R.L.}},
\bauthor{\bsnm{Stenbakken}, \binits{G.N.}},
\bauthor{\bsnm{Williams}, \binits{E.R.}}:
\batitle{Alignment uncertainties of the {NIST} watt experiment}.
\bjtitle{IEEE Trans. Instrum. Meas.}
\bvolume{46},
\bfpage{605}--\blpage{608}
(\byear{1997})
\end{barticle}
\endbibitem

%%% 29
\bibitem{Kramar2002}
\begin{bchapter}
\bauthor{\bsnm{Kramar}, \binits{J.}},
\bauthor{\bsnm{Newell}, \binits{D.}},
\bauthor{\bsnm{Pratt}, \binits{J.}}:
\bctitle{{NIST} electrostatic force balance experiment}.
In: \bbtitle{International Measurement Confederation}
(\byear{2002})
\end{bchapter}
\endbibitem

%%% 30
\bibitem{Cosandier2014}
\begin{barticle}
\bauthor{\bsnm{Cosandier}, \binits{F.}},
\bauthor{\bsnm{Eichenberger}, \binits{A.}},
\bauthor{\bsnm{Baumann}, \binits{H.}},
\bauthor{\bsnm{Jeckelmann}, \binits{B.}},
\bauthor{\bsnm{Bonny}, \binits{M.}},
\bauthor{\bsnm{Chatagny}, \binits{V.}},
\bauthor{\bsnm{Clavel}, \binits{R.}}:
\batitle{Development and integration of high straightness flexure guiding
  mechanisms dedicated to the {METAS} watt balance {Mark II}}.
\bjtitle{Metrologia}
\bvolume{51},
\bfpage{88}--\blpage{95}
(\byear{2014})
\end{barticle}
\endbibitem

\end{thebibliography}
\end{document}